\documentclass[showpacs,twocolumn,superscriptaddress]{revtex4}

\usepackage{doi}
\usepackage{hyperref}
\hypersetup{
  colorlinks=true,        
  linkcolor=blue,         
  citecolor=cyan,         
}

\usepackage{graphicx}
\usepackage{dcolumn}
\usepackage{bm}
\usepackage{color}

\usepackage{amsmath}
\usepackage{amssymb}

\begin{document}

\title{Charged particle motion around non-singular black holes in conformal gravity in the presence of external magnetic field}

\author{Bakhtiyor Narzilloev}
\email[]{nbakhtiyor18@fudan.edu.cn}
\affiliation{Center for Field Theory and Particle Physics and Department of Physics, Fudan University, 200438 Shanghai, China }
\affiliation{Ulugh Beg Astronomical Institute, Astronomicheskaya 33,
	Tashkent 100052, Uzbekistan }

\author{Javlon Rayimbaev}
\email[]{javlon@astrin.uz}
\affiliation{Ulugh Beg Astronomical Institute, Astronomicheskaya 33,
	Tashkent 100052, Uzbekistan }
\affiliation{Institute of Nuclear Physics, Tashkent 100214, Uzbekistan}
\affiliation{National University of Uzbekistan, Tashkent 100174, Uzbekistan}

\author{Ahmadjon Abdujabbarov}
\email[]{ahmadjon@astrin.uz}

\affiliation{Shanghai Astronomical Observatory, 80 Nandan Road, Shanghai 200030, P. R. China}
\affiliation{Ulugh Beg Astronomical Institute, Astronomicheskaya 33,
	Tashkent 100052, Uzbekistan }
\affiliation{Institute of Nuclear Physics, Tashkent 100214, Uzbekistan}
\affiliation{National University of Uzbekistan, Tashkent 100174, Uzbekistan}
\affiliation{Tashkent Institute of Irrigation and Agricultural Mechanization Engineers, Kori Niyoziy, 39, Tashkent 100000, Uzbekistan}

\author{Cosimo Bambi}
\email[]{bambi@fudan.edu.cn}
\affiliation{Center for Field Theory and Particle Physics and Department of Physics, Fudan University, 200438 Shanghai, China }

\date{\today}
\begin{abstract}

We consider electromagnetic fields and charged particle dynamics around non-singular black holes in conformal gravity immersed in an external, asymptotically uniform magnetic field. First, we obtain analytic solutions of the electromagnetic field equation around rotating non-singular black holes in conformal gravity. We show that the radial components of the electric and magnetic fields increase with the increase of the parameters $L$ and $N$ of the black hole solution. Second, we study the dynamics of charged particles. We show that the increase of the values of the parameters $L$ and $N$ and of magnetic field causes a decrease in the radius of the innermost stable circular orbits (ISCO) and the magnetic coupling parameter can mimic the effect of conformal gravity giving the same ISCO radius up to $\omega_{\rm B}\leq 0.07$ when $N=3$. 

\end{abstract}
\maketitle


\section{Introduction}
\label{introduction}

General Relativity has a good success in predicting and describing a number of observations and experiments about gravity. The recent detection of gravitational waves~\cite{LIGO16} and observation of black hole shadow~\cite{EHT19a} can also be considered tests of General Relativity~\cite{EHT19b,LIGO16e}, which successfully passed them. However, the theory itself is plagued by the problem of singularities. Most physically-relevant solutions of Einstein's equation possess a singular region. The current standard understanding of physical laws cannot accept a singularity and there are several attempts to resolve the singularity problem (see, e.g.,~\cite{Bardeen68,Hayward06,Ayon-Beato98,Ayon-Beato99,Ayon-Beato99a,Fan16,Toshmatov14,Englert76,Narlikar77,Mannheim12,Bars14,Bambi13b,Bambi17,Bambi16c,Horava09,Horava09a,Kehagias09}). 

One of the examples to resolve the singularity problem has been proposed in Refs.~\cite{Bambi17}, where, within a large class of conformally invariant theories of gravity, singularity-free black hole solutions have been proposed. The proposed spacetimes are geodetically complete and the curvature invariants do not diverge at $r=0$.

Ref.~\cite{Turimov18b} has been devoted to study the electromagnetic field around compact starts in conformal gravity. Using X-ray observations of supermassive black holes (SMBHs), it was proposed a test of conformal gravity in~\cite{Zhou18}. The energy condition and scalar perturbations of the spacetime in conformal gravity have been studied in~\cite{Toshmatov17b,Toshmatov17a}. Recently, magnetized particle motion around black holes in conformal gravity in the presence of external magnetic fields has been studied in~\cite{Haydarov20}. 
In this work, we plan to investigate charged particle motion acceleration around singularity-free black
holes immersed in an external magnetic field. 

A black hole itself cannot have its own magnetic field. However the curved spacetime will change the structure of electromagnetic field surrounding the black hole. The pioneer work of Wald has been devoted to study the electromagnetic field around a Kerr black hole immersed in an external, asymptotically uniform magnetic field~\cite{Wald74}. A number of works have been devoted to study the electromagnetic field and charged particle motion around compact objects in external magnetic fields~\cite{Aliev86,Aliev89, Aliev02, Frolov11, Frolov12, Benavides-Gallego18,Shaymatov18, Oteev16, Toshmatov15d, Stuchlik14a, Abdujabbarov14,Abdujabbarov10, Abdujabbarov11a, Abdujabbarov11, Abdujabbarov08,Karas12a, Stuchlik16,Kovar10,Kovar14,Kolos17,
Rayimbaev16,Rahimov11a,Rahimov11,Haydarov20b,Rayimbaev20a, Narzilloev19}. 

The energetic process around black holes can be used to model the observational features of astrophysical objects (Relativistic jets, Soft gamma ray repeaters -  SGR and etc.). Different mechanisms of energy extraction from black holes have been proposed: Penrose process~\cite{Penrose69a}, Blandford-Znajeck mechanism~\cite{Blandford1977}, Magnetic Penrose process~\cite{Dhurandhar83,Dhurandhar84,Dhurandhar84b,wagh85}, and particle acceleration mechanism (BSW)~\cite{Banados09}. Particularly, in Ref.~\cite{Banados09} it was shown that the center of mass energy of colliding particles near an extreme rotating Kerr black hole may diverge for the fine tuned values of the angular momentum of the particles. Magnetic fields may play an important role in the charged particle acceleration near black holes~\cite{Frolov11,Frolov12}. Thus we are also interested to study the effects of conformal gravity and magnetic fields on the charged particle acceleration process.

In this paper, we study the dynamics of charged particles around black holes in conformal gravity in the presence of magnetic fields. The paper is organized as follows: Sect.~\ref{sec2} is devoted to study the electromagnetic field around black holes in conformal gravity immersed in an external, asymptotically uniform magnetic field. The charged particle motion around non-rotating and rotating black holes in conformal gravity in the presence of an external magnetic field has been studied in Sects.~\ref{sec_2} and \ref{sec_3}, respectively. In Sect.~\ref{Sec:collision}, we study the collision of charged particles near a black hole in conformal gravity. We summarize our results in Sect.~\ref{Sec:Conclusion}. 
Throughout this work, we use $(-,+,+,+)$ 
for the space-time signature and a system of units where
$G = c = 1$ . 

\section{Electromagnetic field around a singularity free Black hole \label{sec2}}

In this section, we study the electromagnetic field around a singularity-free black hole in conformal gravity described by the line element
\begin{eqnarray}\label{metric1}
ds^{*2}=S(r,\theta)ds^2_{Kerr}\ ,
\end{eqnarray}
where
\begin{equation}
ds^2_{Kerr} = g_{\alpha \beta}^{K}dx^\alpha dx^\beta\ , \label{metric}
\end{equation}
with
\begin{eqnarray}
g_{00}^{K}&=&-\left(1-\frac{2 M r}{\Sigma}\right)\ , \nonumber \\
 g_{11}^{K}&=&\frac{\Sigma}{\Delta}\ , \nonumber \\
g_{22}^{K}&=&\Sigma\ ,  \\
 g_{33}^{K}&=&\left[(r^2+a^2)+\frac{2 a^2 M r \sin^2
\theta}{\Sigma}\right]\sin^2
\theta\ , \nonumber \\\nonumber
 g_{03}^{K}&=&-\frac{2 M a r \sin^2 \theta}{\Sigma}\ ,\label{1}
\end{eqnarray}
\begin{equation}
\Delta = r^2 + a^2 -2 M r , \;\;\;\;\; \Sigma=r^2+a^2 \cos^2\theta
\ , \nonumber \label{p2}
\end{equation}
and
\begin{eqnarray}
S(r,\theta)= \left(1+\frac{L^2}{\Sigma}\right)^{2N+2}\ ,
\end{eqnarray}
where $a$ is the spin parameter of the black hole with total mass $M$, {$S$ is called the conformal rescaling,} and $L$ and $N$ are conformal parameters; that is, parameters related to the conformal rescaling.

{The Penrose diagram, and thus the causal structure of the spacetime, of our singularity-free black hole solution is the same as the Penrose diagram of the Kerr solution, because conformal transformations do not alter the causal structure of spacetimes. They only change distances. The Kerr singularity at $r=0$ is not a curvature singularity any longer in our solution, and this is possible because the scalar curvature and the Kretschmann scalar are not invariant under conformal trasformations. After the conformal rescaling, massive particles need an infinite time to reach the point $r=0$. Massless particles require an infinite value of their affine parameter to reach the point $r=0$. So the point $r=0$ is not a singularity in the sense of geodesic motion any longer either. The conformal rescaling $S$ has thus the capability to remove the Kerr singularity at $r=0$. $S$ is also singular at $r=0$ and this is strictly necessary to compensate the singularity of the Kerr solution at $r=0$. For $N \ge 1$, the conformal factor can remove the singularity of the Kerr metric at $r=0$, while, for a lower value, the singularity of $S$ is not strong enough to compensate the singularity of the Kerr metric. More details can be found in the original paper deriving this solution~\cite{Bambi17}}

%
%
We start with considering the electromagnetic field around a singularity-free black hole immersed in an external, asymptotically uniform magnetic field aligned along the direction of the axis of
symmetry of the space-time.
The energy momentum tensor of the electromagnetic field  is assumed to be negligibly small, does not change the spacetime metric, and is of the following order of magnitude
\begin{eqnarray}
&& \frac{B^2 r^3}{8\pi M c^2}\simeq \frac{3}{400} 
\left(\frac{B}{10^{3}\ {\rm G}}\right)^2 \left(\frac{
M_{\bigodot}}{M}\right) \left(\frac{r}{1.5\ {\rm km}}\right)^3 .
\end{eqnarray}

Using a Killing vector $\xi^\mu$ being responsible for the symmetry of spacetime geometry of the black hole, one may recall the equation
\begin{equation}\label{2.1}
 \xi_{\alpha ;\beta}+\xi_{\beta;\alpha}=0\ .
\end{equation}
Expressions (\ref{2.1}) can be used to have the following equation
\begin{equation}
\xi_{\alpha;\beta;\gamma}-\xi_{\alpha;\gamma;\beta}=-\xi^\lambda
R_{\lambda\alpha\beta\gamma}\ .
\end{equation}
where the Riemann curvature tensor $R_{\lambda\alpha\beta\gamma}$ can be transformed to the Ricci one in the following way
\begin{equation}\label{ricci}
\xi^{\alpha;\beta}_{\ \ \ ;\beta}=\xi^\gamma R_{\gamma\beta}^{\ \
\ \alpha\beta}=R^{\alpha}_{\ \gamma} \xi^{\gamma}\ .
\end{equation}

For the spacetime of the singularity free rotating black hole in the conformal gravity, the right hand side of Eq.~(\ref{ricci}) can
be chosen as $R^{\alpha}_{\ \gamma} \xi^{\gamma}=\eta^{\alpha}$. The Maxwell equations then can be expressed in the following form
\begin{equation}\label{waldself}
F^{\alpha\beta} _{\ \ \ ;\beta}=-2\xi^{\alpha;\beta} _{\ \ \
;\beta}+2 \eta^{\alpha}=0\ ,
\end{equation}
where $F_{\alpha\beta}$ is electromagnetic field tensor and can be selected as
\begin{equation}
F_{\alpha\beta}=C_0 \xi_{[\beta;\alpha]}+f_{\alpha\beta}=-2C_0\left(\xi_{\alpha;\beta}\right)+a_{[\beta,\alpha]}\ .
\end{equation}
Here $C_0$ is an integration constant and the 4-vector $a^\alpha$ is a correction to the potential due to the presence of the conformal gravity parameters responsible for the Ricci non flatness of the spacetime. The  4-vector $a^\alpha$ can be found from the equation $ \Box{a^\alpha}=\eta^{\alpha}$.

The electromagnetic potential then can be written as the sum of two contributions
\begin{equation}
A^\alpha=\tilde{A}^\alpha+ a^\alpha.
\end{equation}
where $\tilde{A}^\alpha$ is the potential being
proportional to the Killing vectors. To find the solution for
$\tilde{A}^\alpha $, one can use the ansatz for the vector potential
$\tilde{A}_\alpha$ of the electromagnetic field in the Lorentz
gauge in the form
$
 \tilde{A}^\alpha=C_1 \xi^\alpha_{(t)}+C_2
\xi^\alpha_{(\varphi)}\ $ \cite{Wald74}.
The constant $C_2=B/2$, where the gravitational source is immersed in
a uniform magnetic field $\textbf{B}$ being parallel to its axis
of rotation. The value of the remaining constant $C_1=aB$ can be
easily calculated from the asymptotic properties of the spacetime
(\ref{metric}) at the asymptotical infinity.

The second part $a^\alpha$ of the total vector
potential of the electromagnetic field is produced due to the contribution of conformal gravity and has the following solution $a^\alpha=\left\{
k B L^2/r,0,0,0\right\}$, where the expression for the constant $k
$ can be easily found from the asymptotic properties of
the spacetime (\ref{metric}) at infinity~\cite{Abdujabbarov08,Abdujabbarov10}. However, since the effect of conformal gravity is negligibly small at large distances, one might exclude this part tending $k\rightarrow0$ and use the traditional contravariant expression for the 4-vector potential as in the Kerr case.

Finally, the components of the 4-vector potential $A_\alpha$ of the electromagnetic field will take a form
\begin{eqnarray}
A_0&=&\frac{a B \left(L^2+\Sigma \right)^4 \left( 2 M r-\Sigma-M r \sin ^2\theta \right)}{\Sigma
   ^5}  \ ,\ \\
A_1&=&0 \ ,
\\
A_2&=&0 \ ,
\\
\label{potentials} A_3 &=& \frac{B \sin ^2\theta  \left(L^2+\Sigma \right)^4 }{2 \Sigma ^5}\\\nonumber
&\times& \left(2 a^2 M r \sin ^2\theta +a^2
   (\Sigma -4 M r)+r^2 \Sigma \right)  ,
\end{eqnarray}
where, for simplicity, we take $N=1$. 
The components of the electric and magnetic fields in the frame moving with four velocity $u_\alpha$ read

\begin{eqnarray}\label{fields}
E_{\alpha} &=& F_{\alpha \beta} u^{\beta}\ ,
\\
B^{\alpha} &=& \frac{1}{2} \eta^{\alpha \beta \sigma \mu} F_{\beta \sigma} u_{\mu}\ ,
\end{eqnarray}
where the electromagnetic field tensor, $F_{\alpha \beta}$, in terms of the four potential can be expressed as
\begin{eqnarray}\label{37}
F_{\alpha \beta}&=& A_{\beta, \alpha}-A_{\alpha, \beta}\ .
\end{eqnarray}

Using the the expressions above for the zero angular momentum observers (ZAMO) with the four-velocity
components

\begin{eqnarray}\label{zamo}
(u^{\alpha})_{_{\rm ZAMO}}&\equiv&
    \sqrt{-g^{00}}\ \left(1,0,0, \frac{g^{03}}{g^{00}}\right) \ ,\\
(u_{\alpha})_{_{\rm ZAMO}}&\equiv&
    {\frac{-1}{\sqrt{-g^{00}}} }\ \big(1,0,0,0 \big)\ , \nonumber
\end{eqnarray}
one can easily find the components of the electromagnetic field using the expressions in (\ref{fields})-(\ref{37}). The nonvanishing orthonormal components of the electromagnetic field measured
by zero angular momentum observers with the four-velocity
that has the form (\ref{zamo}) are given by the expressions in (\ref{er})-(\ref{bt}) and linear approximation of conformal gravity parameter $L^2$ is presented by the expressions in (\ref{e1})-(\ref{m2}) in appendix \ref{app}.

Fig.~\ref{fig_1} shows the radial dependence of the components of electromagnetic fields in the cases of different values of the angle $\theta$ and conformal gravity parameter $L^2$. Since the expressions in (\ref{er})-(\ref{bt}) have a complex form, one might be interesting  to see the structure of the electromagnetic fields around BHs in the ZAMO frame, which is presented in Fig.~\ref{fig1.2}. %

\begin{figure*}[ht!]
\begin{center}
\includegraphics[width=0.31\linewidth]{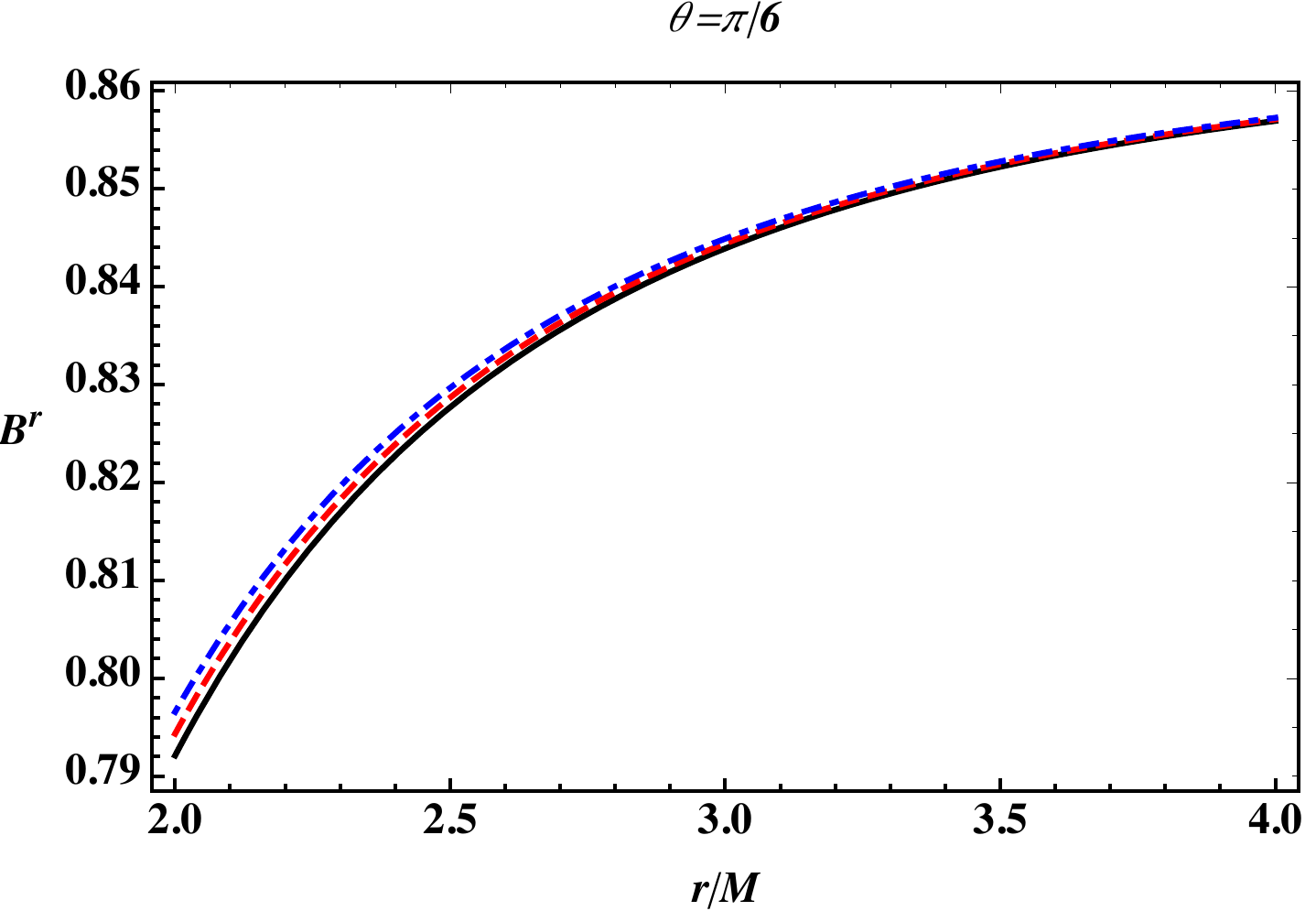}
\includegraphics[width=0.325\linewidth]{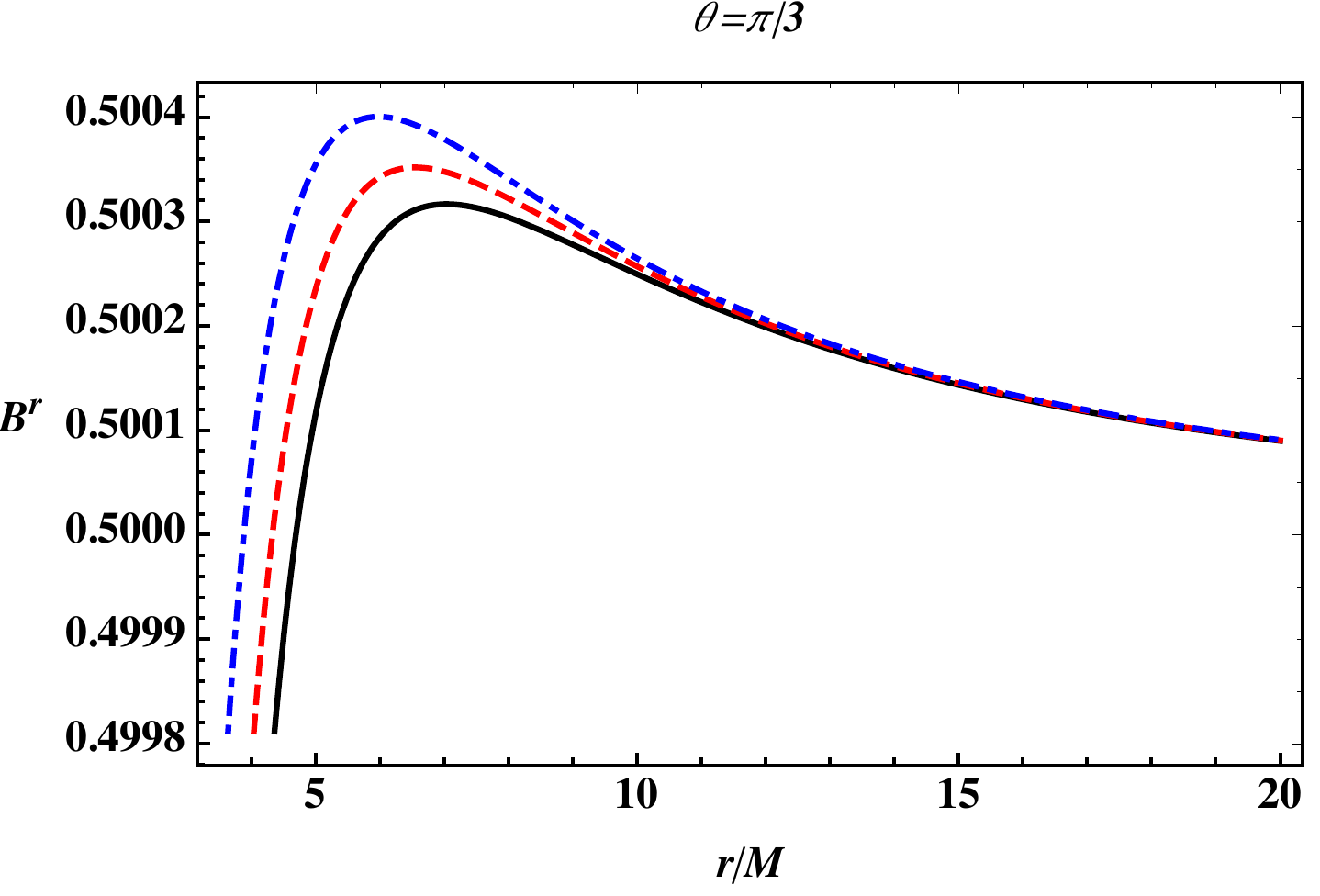}
\includegraphics[width=0.32\linewidth]{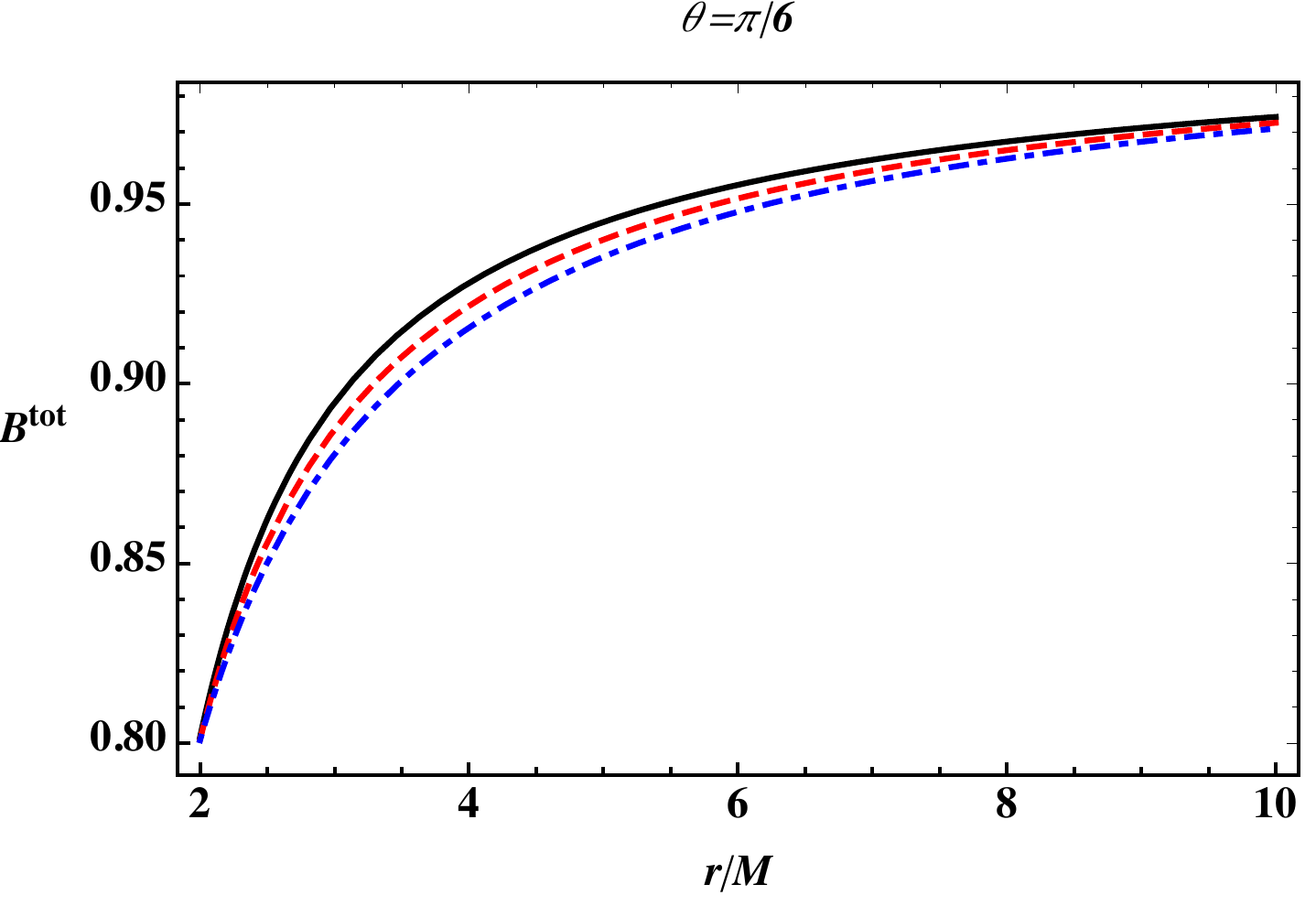}

\includegraphics[width=0.32\linewidth]{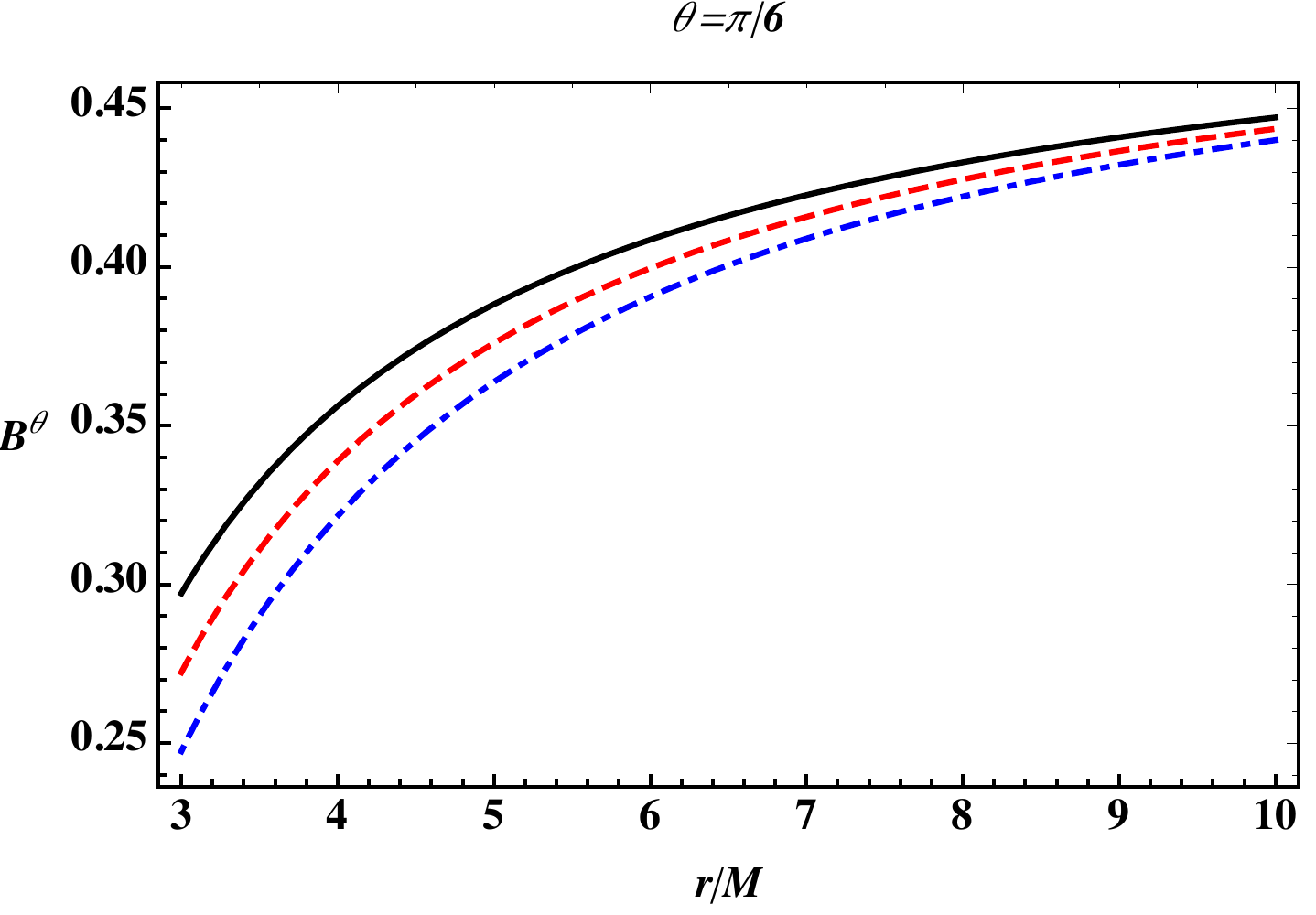}
\includegraphics[width=0.31\linewidth]{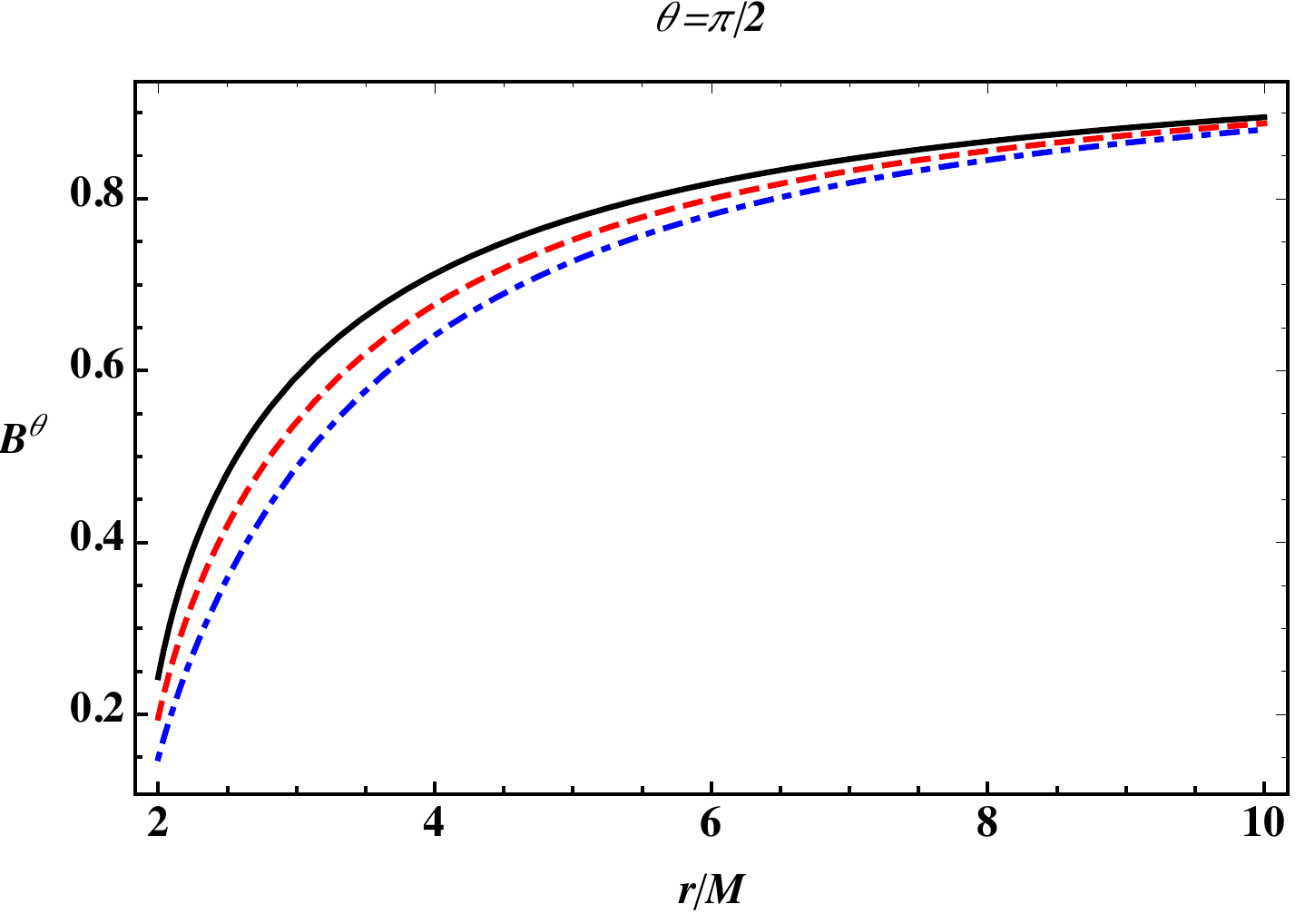}
\includegraphics[width=0.32\linewidth]{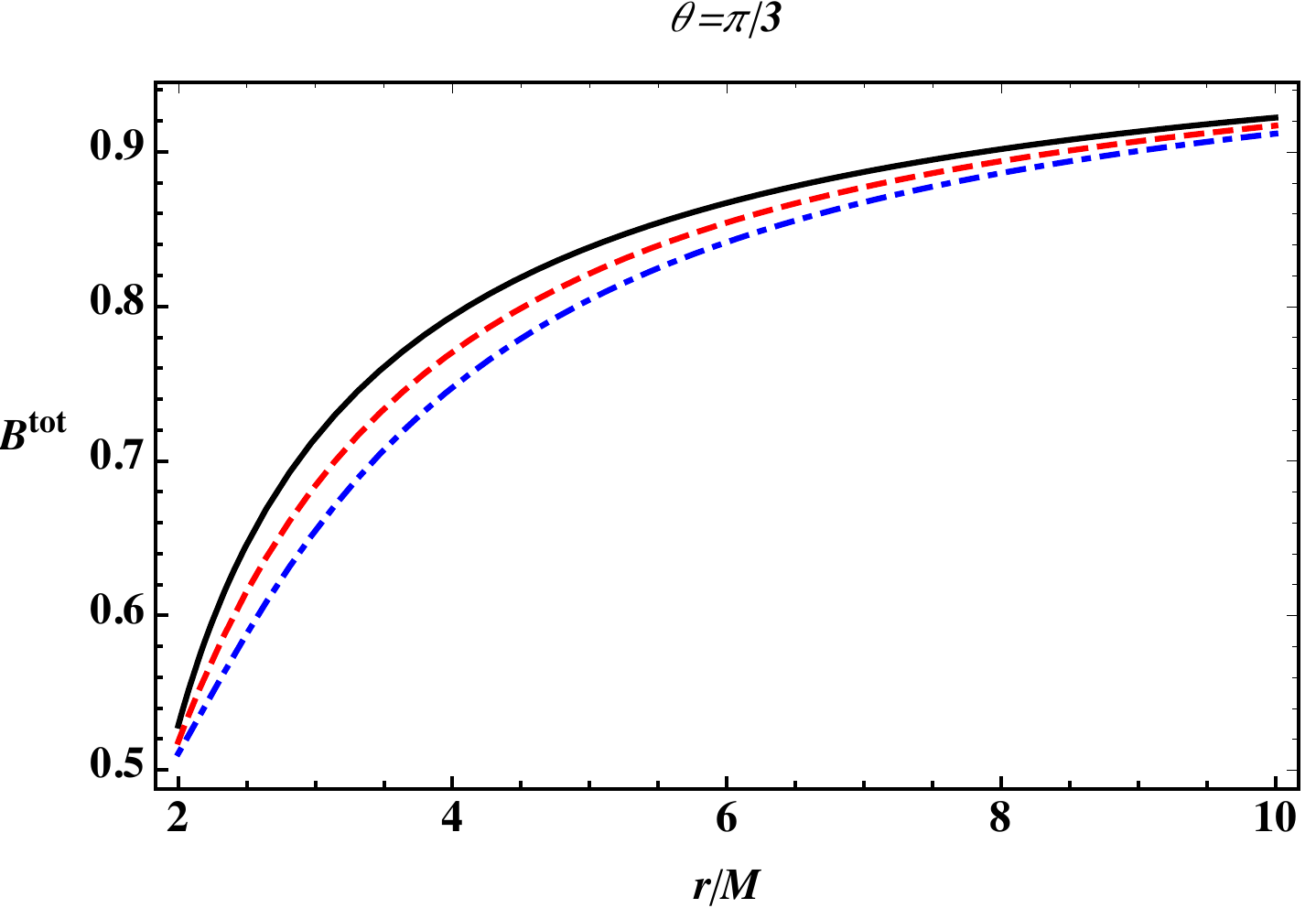}

\includegraphics[width=0.32\linewidth]{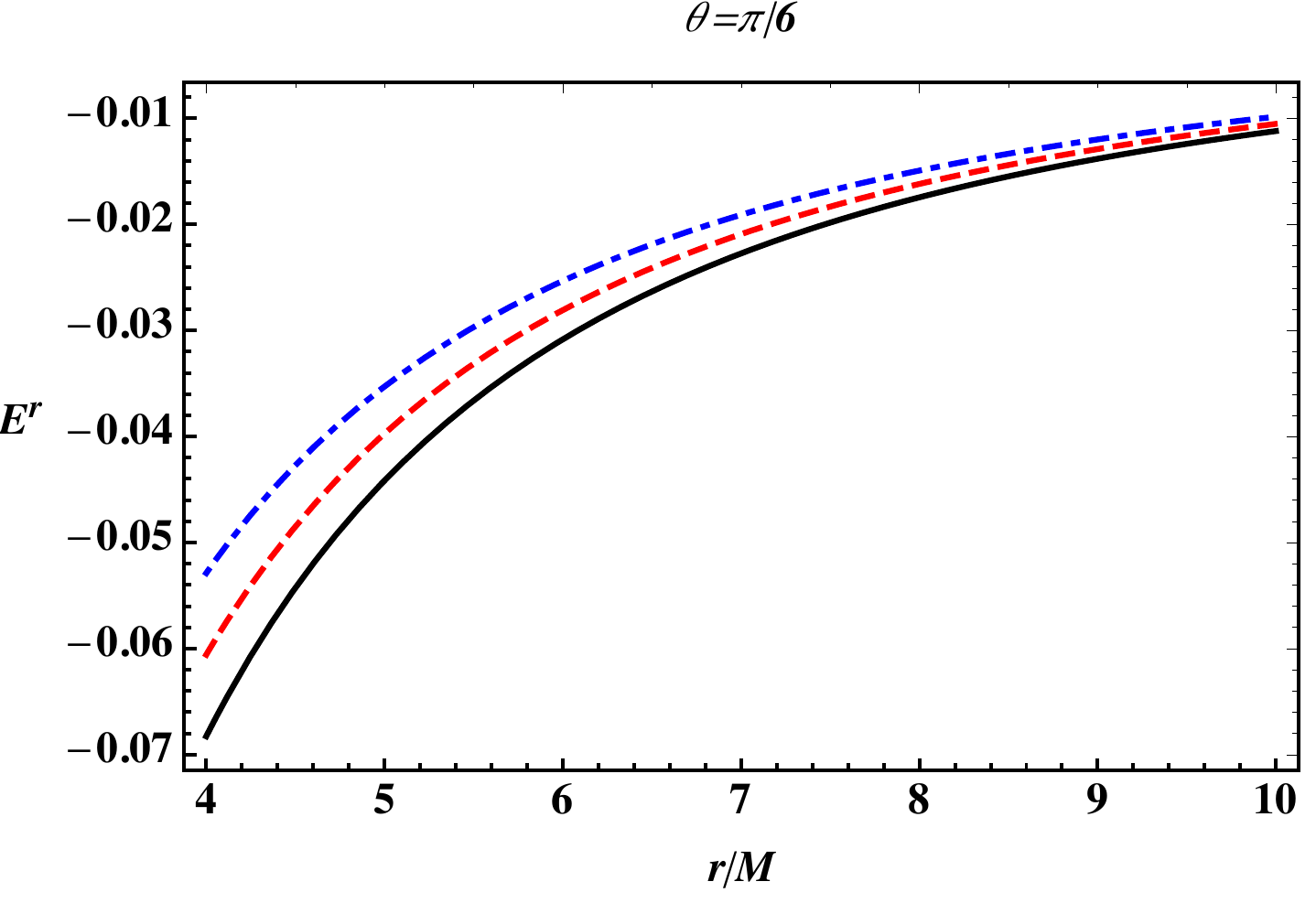}
\includegraphics[width=0.32\linewidth]{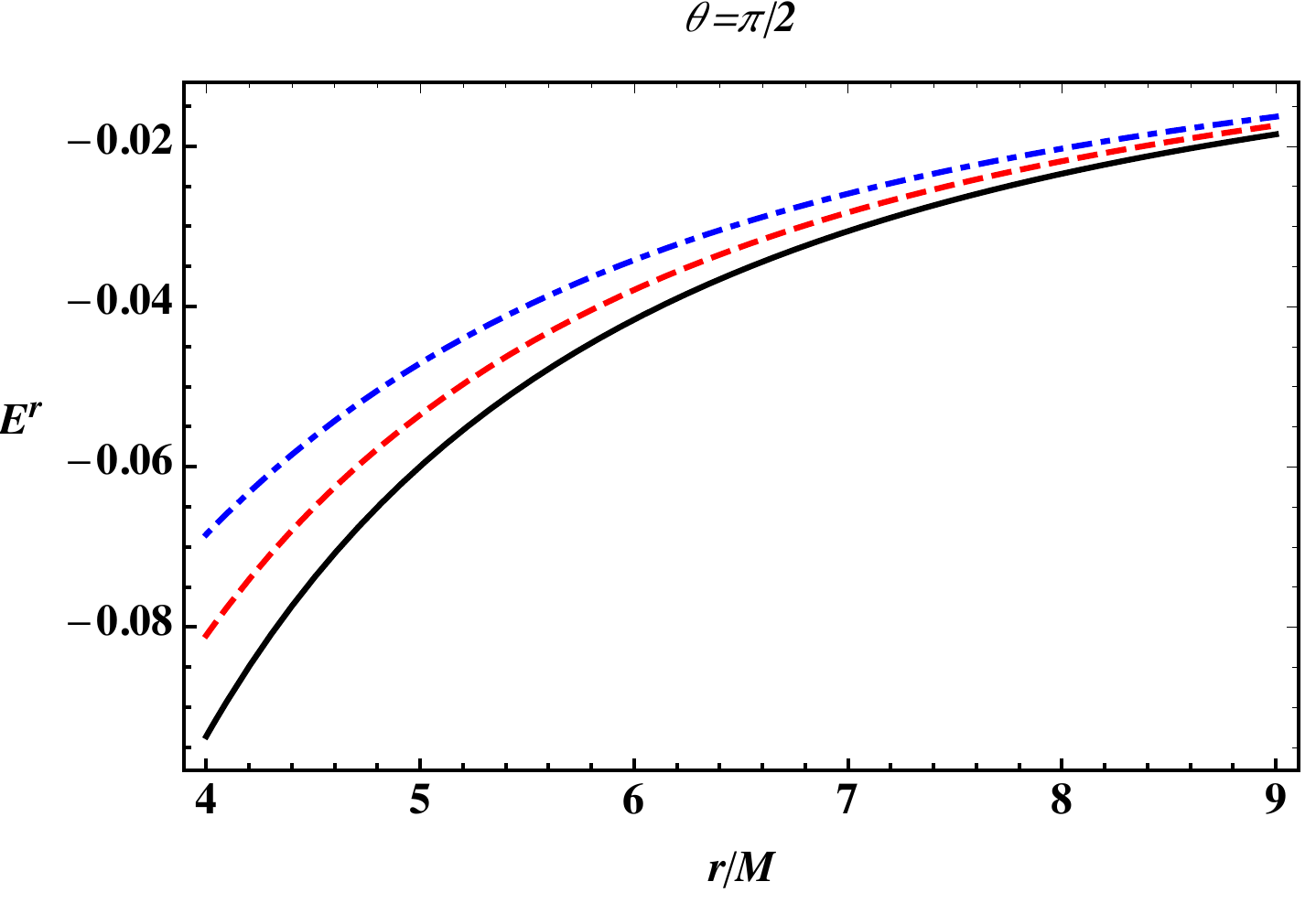}
\includegraphics[width=0.32\linewidth]{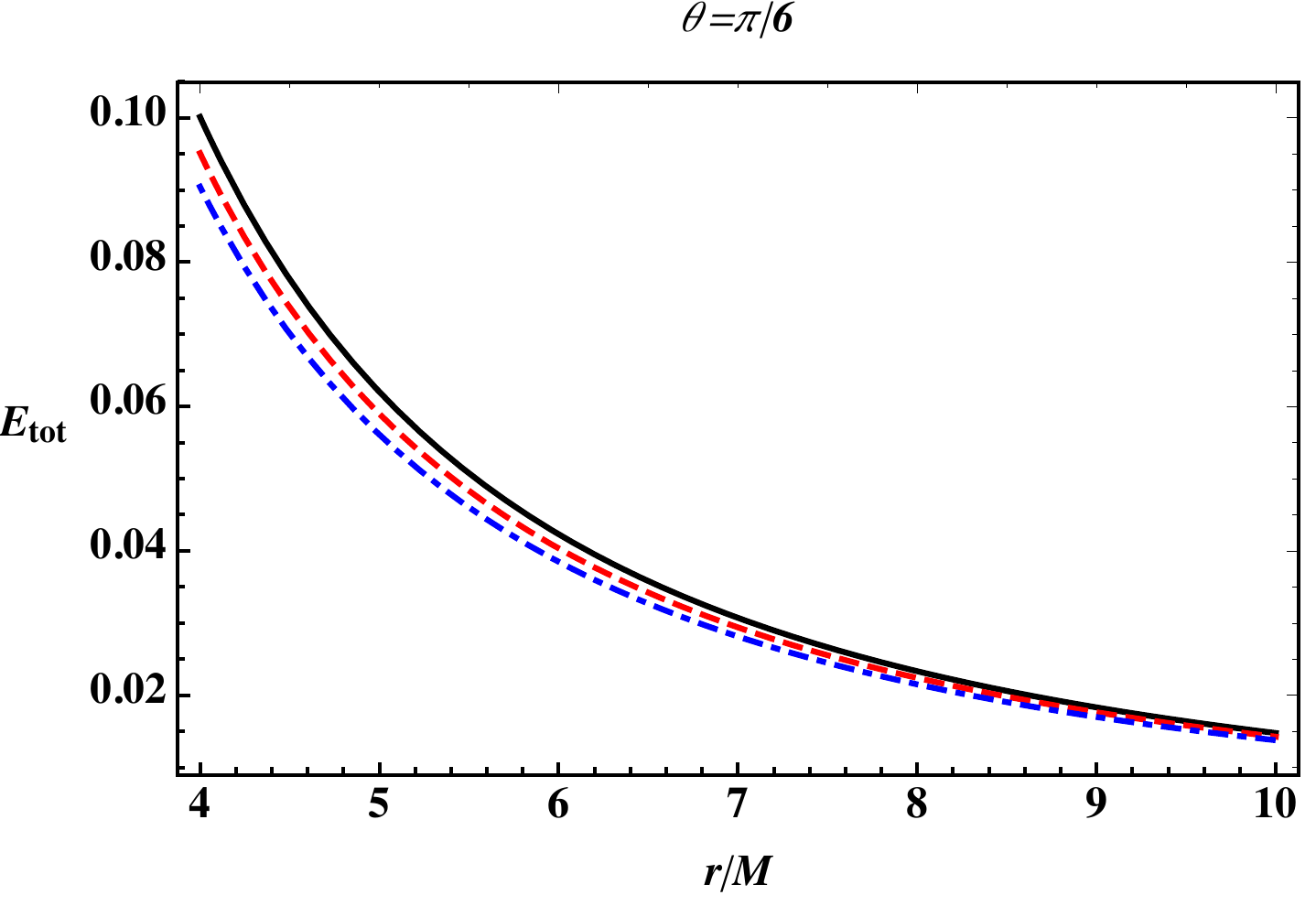}

\includegraphics[width=0.32\linewidth]{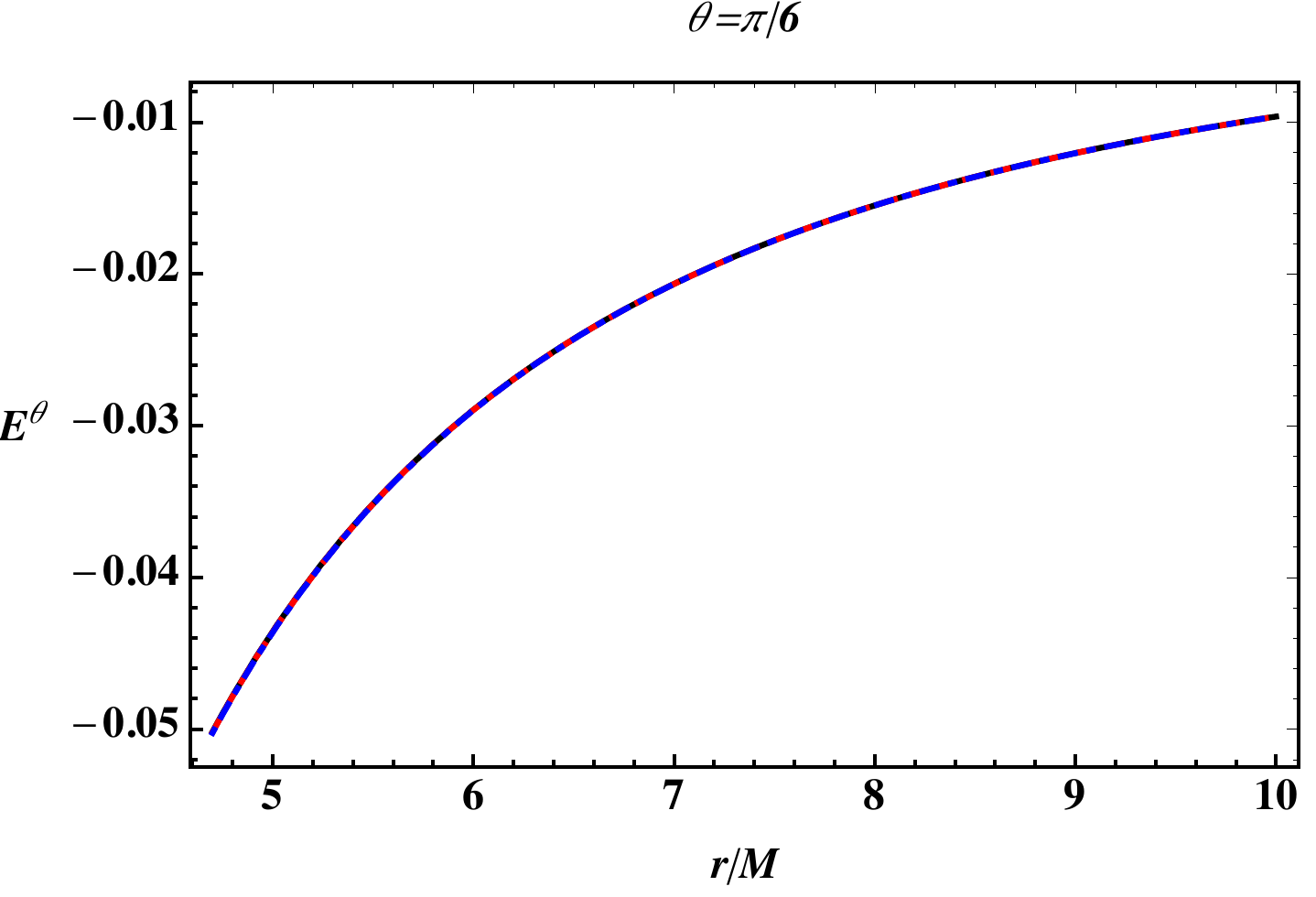}
\includegraphics[width=0.32\linewidth]{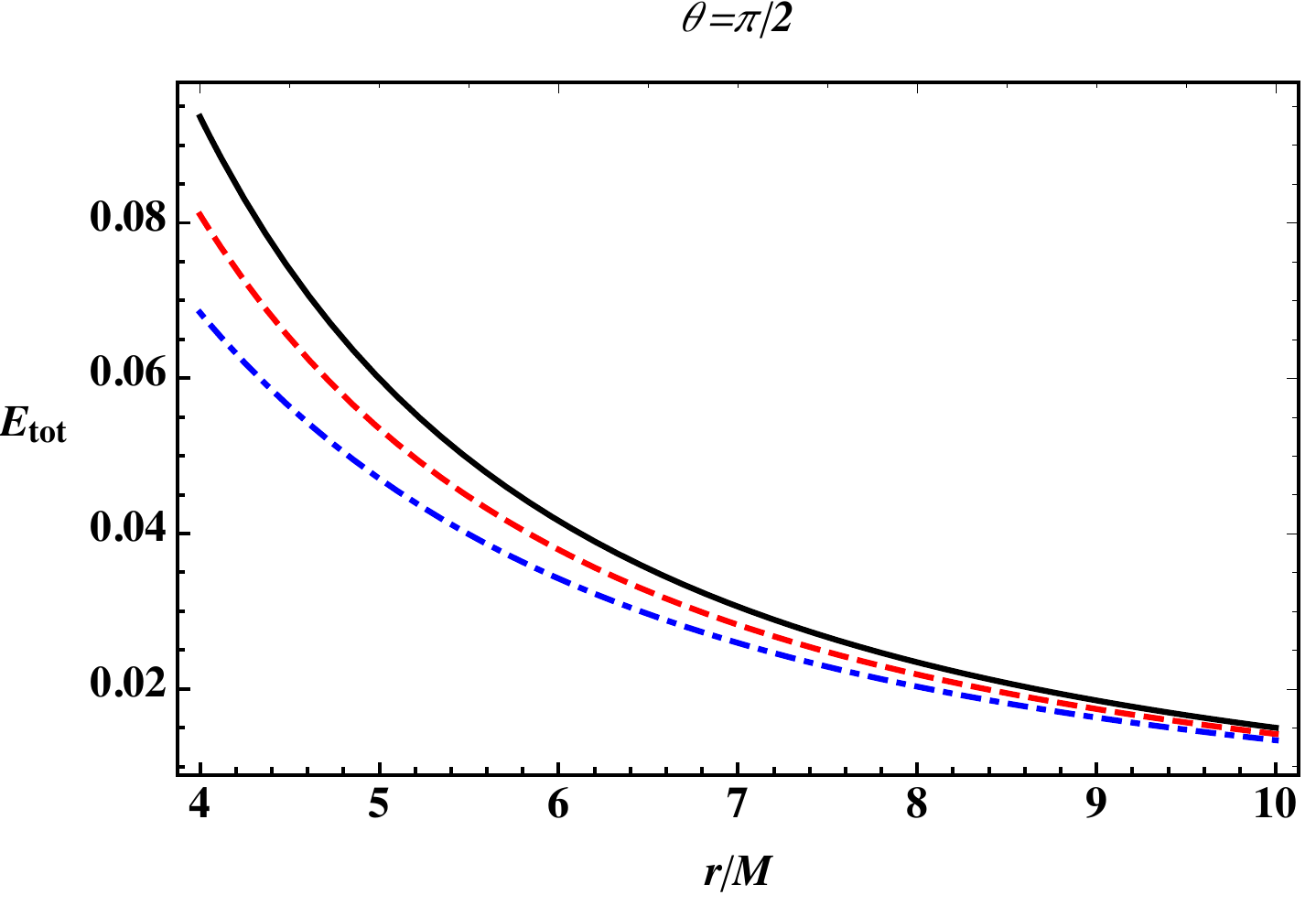}
\includegraphics[width=0.32\linewidth]{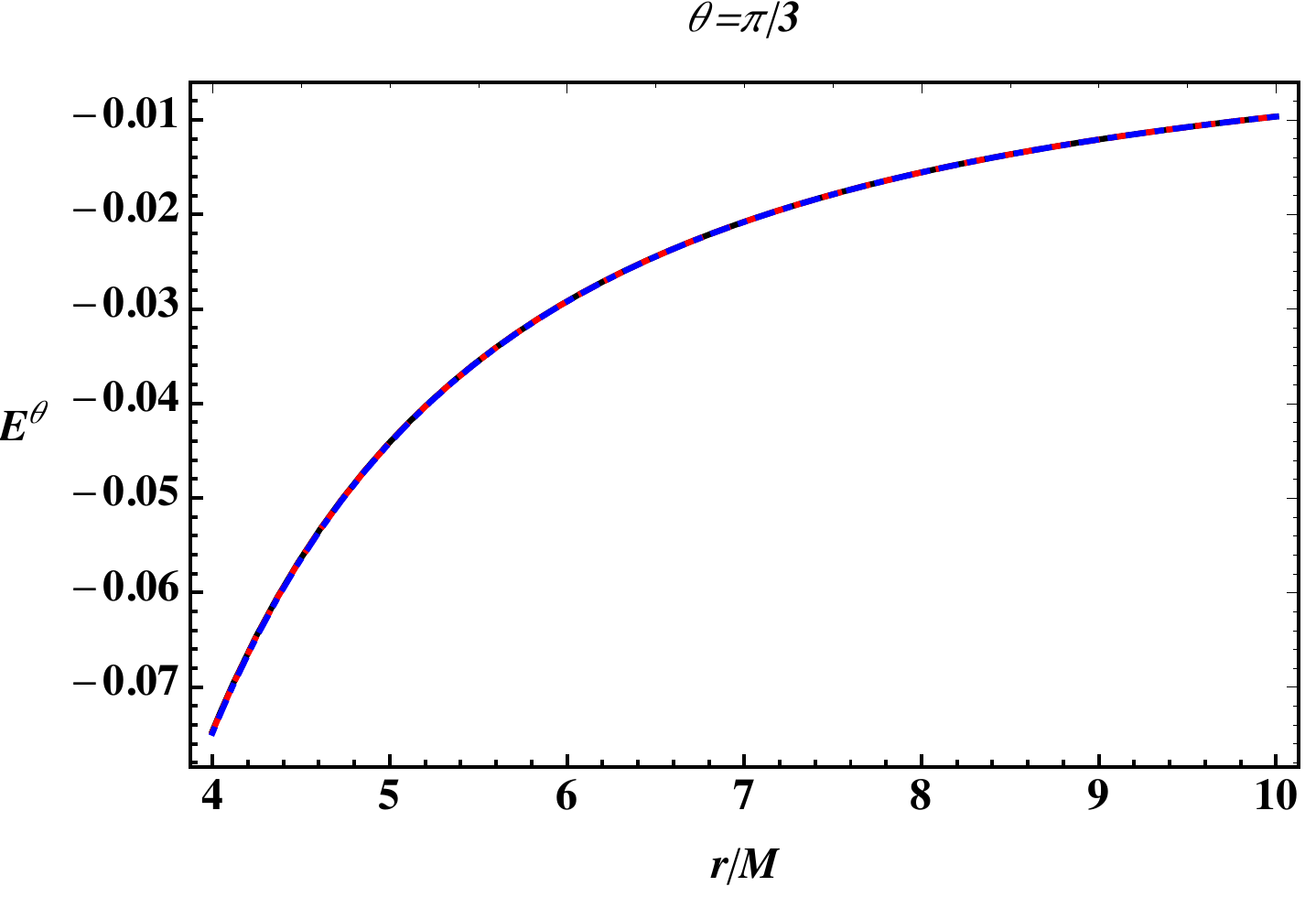}
\end{center}
\caption{Radial dependence of the electric and magnetic fields for different values of $\theta$ . All graphics are plotted in the case of $B=1$ and $M=1$. Black solid lines correspond to $L^2=0$, while dashed red lines and blue dot-dashed lines to $L^2=0.2$ and $L^2=0.4$, respectively.
\label{fig_1}}
\end{figure*}

\begin{figure*}[ht!]
\begin{center}
a.
\includegraphics[width=0.45\linewidth]{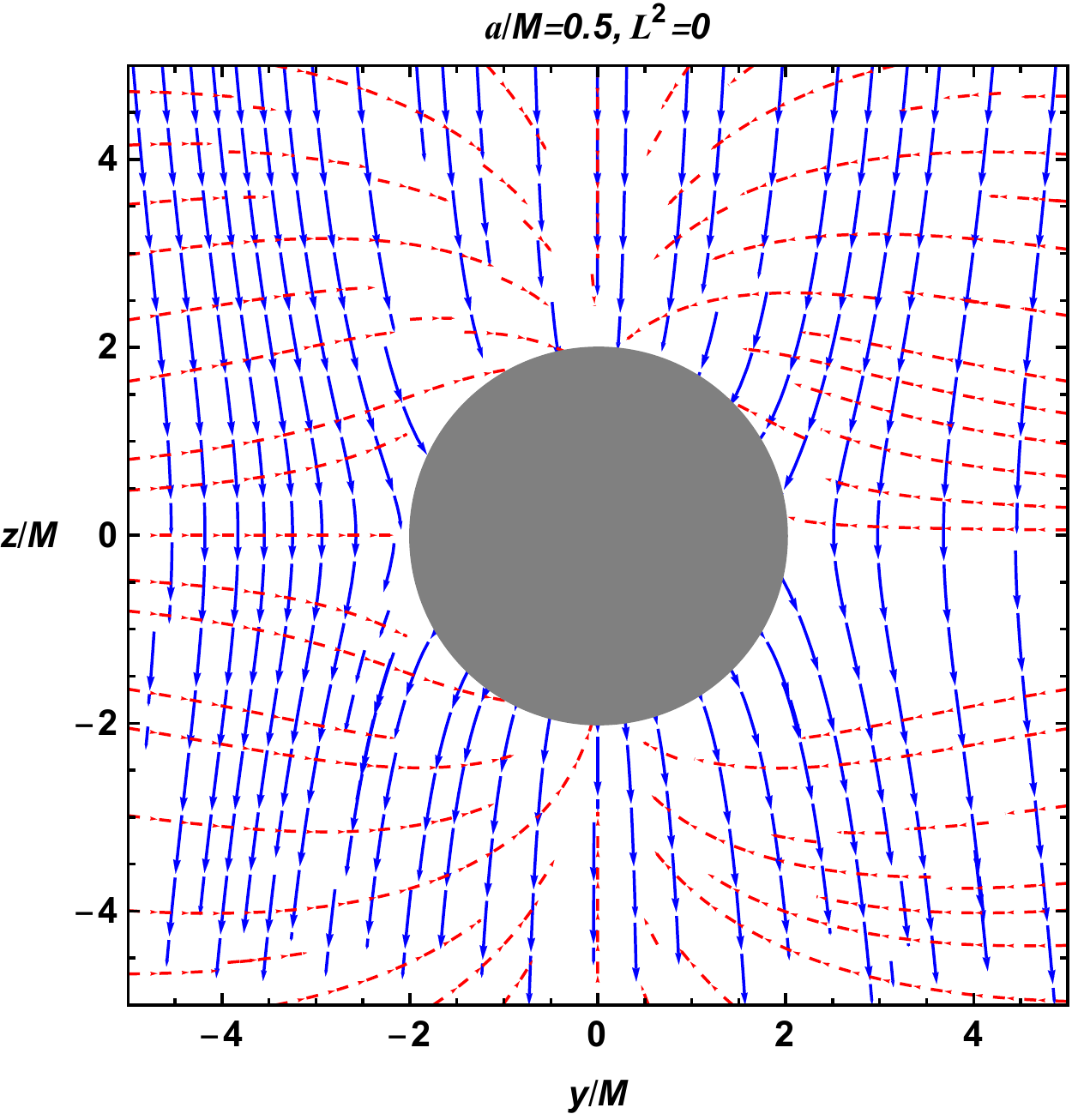}
b.
\includegraphics[width=0.45\linewidth]{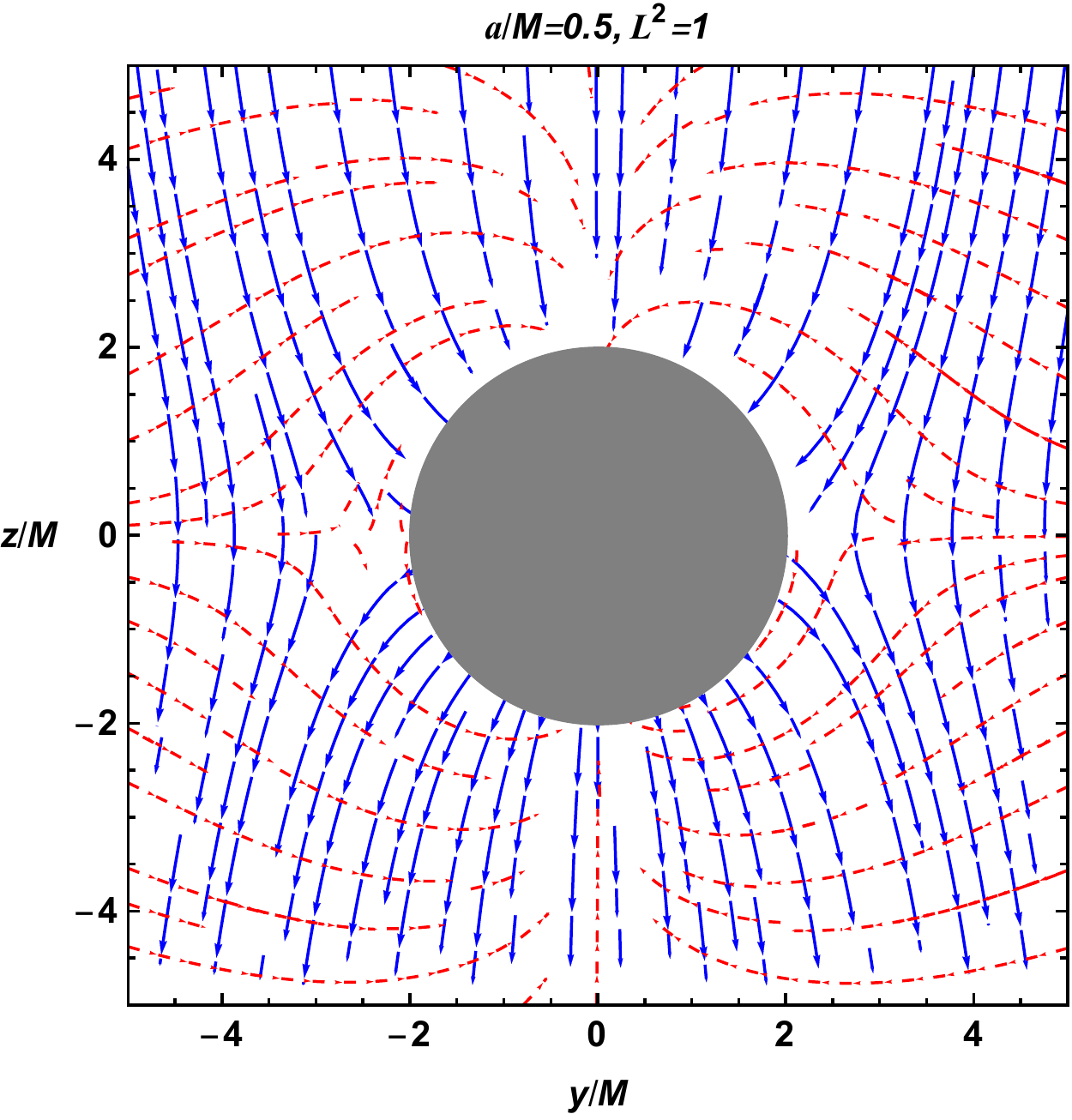}
\end{center}
\caption{Structure of the magnetic and electric fields around a black hole. Blue and red lines correspond to magnetic and electric fields, respectively. \label{fig1.2}}
\end{figure*}

In the case of slow rotation ($a\ll M$) and far distance ($M/r\ll1$), expressions (\ref{er})-(\ref{bt}) reduce to
\begin{eqnarray}
\label{weaker}E^{\hat{r}}&=& -\frac{a B}{r^2} \left[\frac{8 L^2}{r}+M \left(2+\sin ^2\theta \right)\right]\ ,
\\\nonumber
\\
E^{\hat{\theta}}&=& -\frac{2 a B M}{r^2} \left[1+\frac{M}{r}\right] \sin 2 \theta \ ,
\\\nonumber
\\
B^{\hat{r}}&=& B \cos \theta\ ,
\\\nonumber
\\
\label{weakbt}B^{\hat{\theta}}&=& B \sin \theta 
\\\nonumber
&\times& \left[1-\frac{M}{r}-\frac{8 L^2+M^2}{2 r^2}-\frac{M^3-8 M L^2}{2 r^3}\right]\ .
\end{eqnarray}

From the expressions in (\ref{weaker})-(\ref{weakbt}), one can see that in such limits only the components $E^{\hat{r}}$ and $B^{\hat{\theta}}$ are affected by conformal gravity while other components ($E^{\hat{\theta}}$ and $B^{\hat{r}}$) do not include any contribution from $L^2$ (for the chosen approximation). The latter is especially clearly seen for  $E^{\hat{\theta}}$ in the graphs presented in Fig.~\ref{fig_1}. In the
limit of flat spacetime, i.e. for $M/r\rightarrow 0$, $Ma/r^2\rightarrow 0$ and $L^2/r^2\rightarrow 0$, expressions
(\ref{er})--(\ref{bt}) give the following limiting expressions: $B^{\hat r}=B\cos\theta, \ B^{\hat\theta}=B\sin\theta, \ E^{\hat
r}=E^{\hat\theta}=0$, being consistent with the solutions for the
homogeneous magnetic field in Newtonian limit.
\begin{figure*}[ht!]
  \centering
    \includegraphics[width=0.45\linewidth]{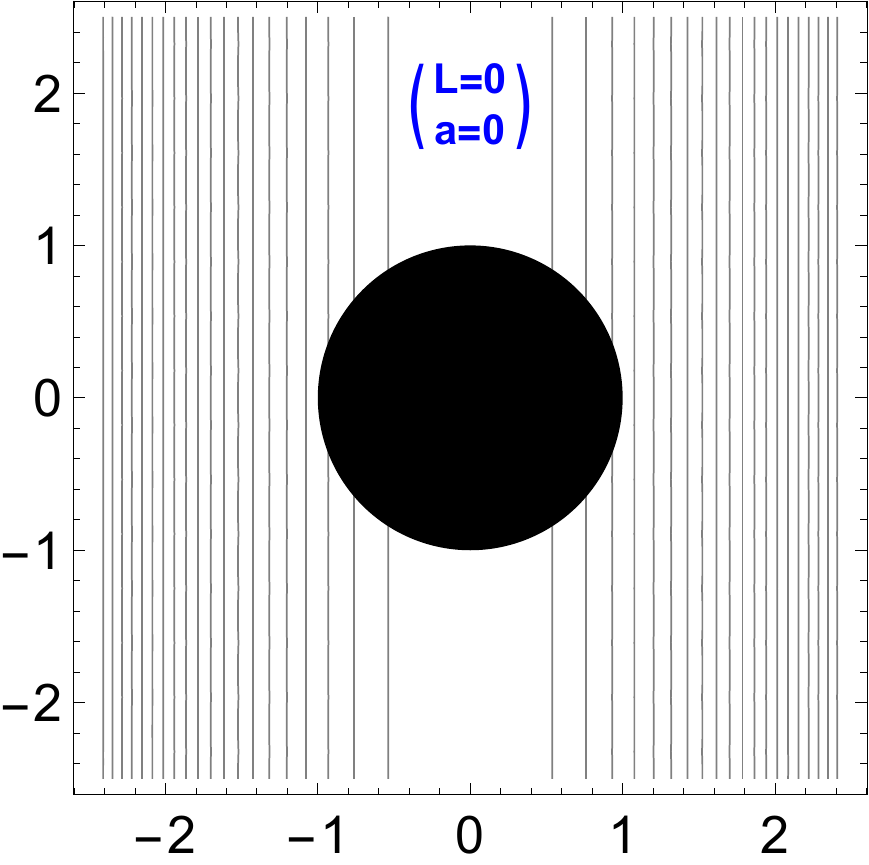}
    \includegraphics[width=0.45\linewidth]{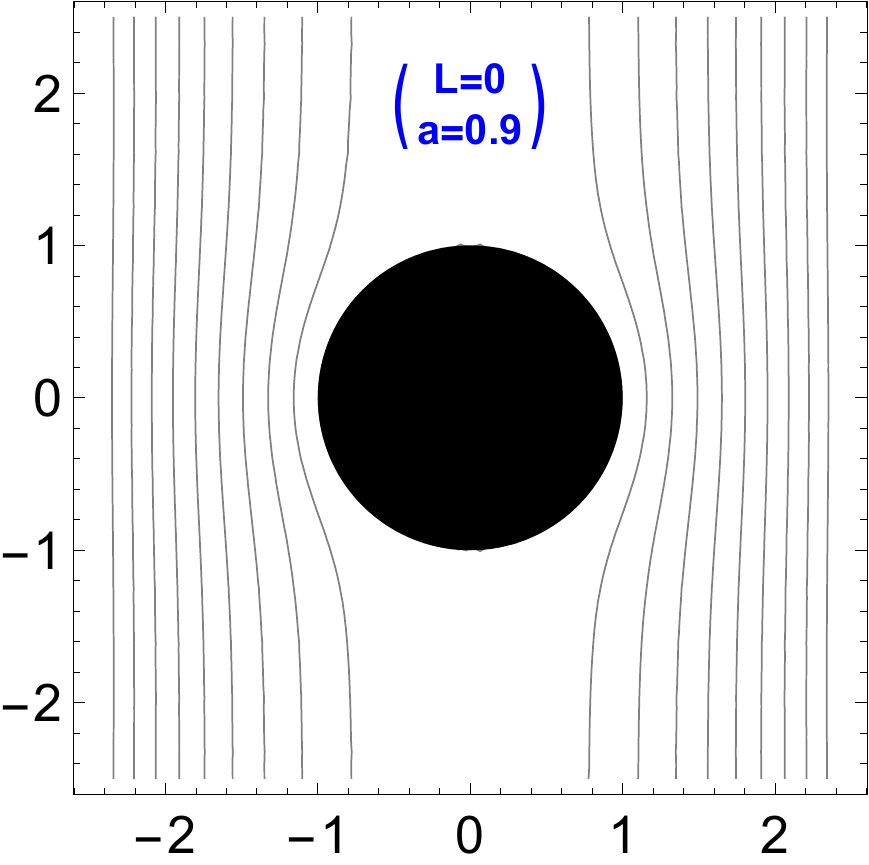}
    \includegraphics[width=0.45\linewidth]{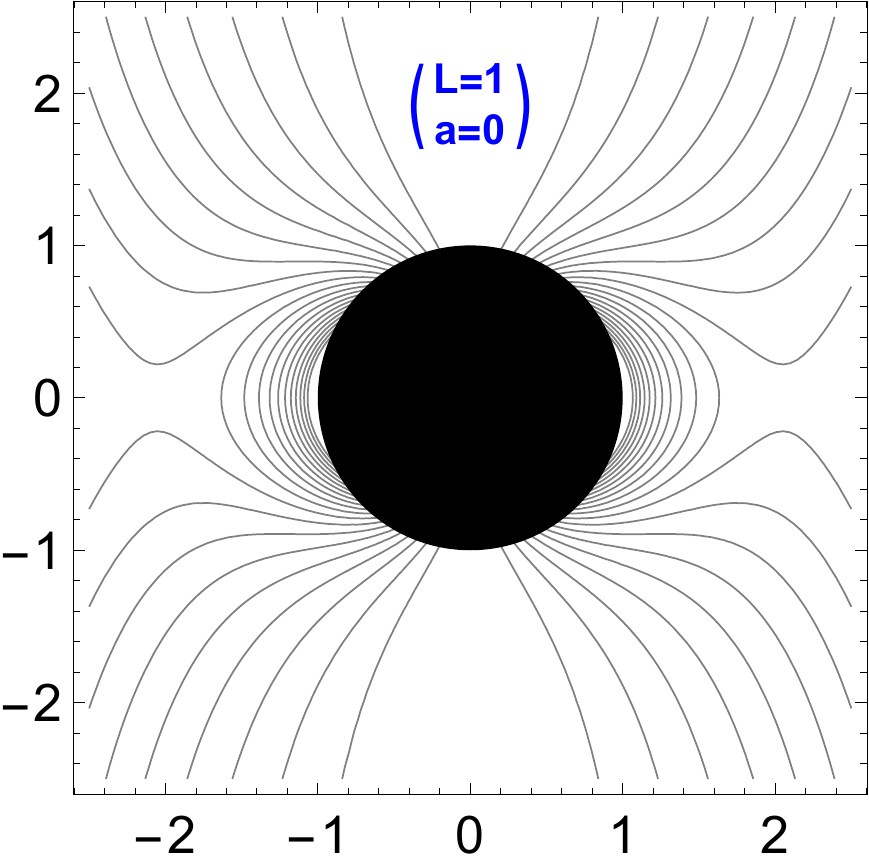}
    \includegraphics[width=0.45\linewidth]{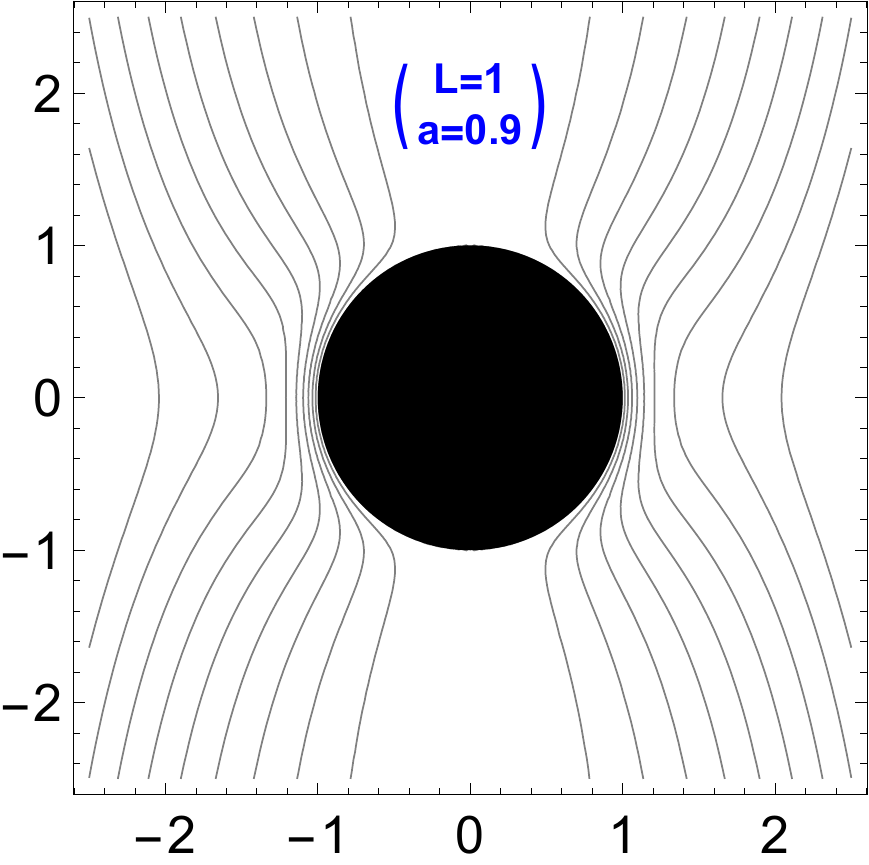}
           \caption{Magnetic field line profiles around a black hole in the $x-z$ plane for different values of spin parameter $a$ and conformal parameter $L$. Units in which $M=1$.}
    \label{mflines}
\end{figure*}
Figure~\ref{mflines} presents the profiles of magnetic fields around black holes for different values of spin and conformal parameter. One can see from the right top and bottom panels that the conformal parameter forces the parallel field line to have a dipole-like structure. Finally, bottom panels give the possibilities of the comparison of the nature of conformal and spin parameters of the black hole.

\section{Charged particle motion around a Conformal non-rotating black hole immersed in a uniform magnetic field\label{sec_2}}

Here we study charged particle motion around a black hole in
conformal gravity in the presence of an external, uniform magnetic field.
The Hamilton-Jacobi equation for a test particle with mass $m$ and charge $q$ can be written as
\begin{eqnarray}\label{HJ}
g^{\mu \nu}\Big(\frac{\partial {\cal S}}{\partial x^{\mu}}-qA_{\mu}\Big)
\Big(\frac{\partial {\cal S}}{\partial x^{\nu}}-q A_{\nu}\Big)=-m^2\ .
\end{eqnarray}
The solution of equation (\ref{HJ}) can be
sought in the following form
\begin{eqnarray}\label{action}
{\cal S}=-E t+l \phi+{\cal S}_r(r)+{\cal S}_{\theta}(\theta)\ ,
\end{eqnarray}
where ${\cal E}=E/m$ and ${\cal L}=l/m$ are the specific energy and the specific angular momentum of the test particle, respectively.

It is convenient to consider the particle motion on the equatorial plane plane, where $\dot{\theta}=0$ ($p_{\theta}=0$). Then we can write 
\begin{eqnarray}
{\dot r}^2 + V_{\rm eff}(r) = {\cal E}^2\, ,
\end{eqnarray}
where the effective potential has a form
\begin{eqnarray}\label{Effpot}
V_{\rm eff}(r)=f(r)\Big[S(r)+\Big(\frac{\cal L}{r}-\omega_{\rm B} r S(r)\Big)^2\Big]
\end{eqnarray}
and $\omega_{\rm B} = eB/2m$ is the magnetic coupling parameter, or the so-called cyclotron frequency, which characterizes the interaction between the charged particle and the magnetic field. The effective potential is invariant under the following transformations: $({\cal L},-\omega_{\rm B}) \longleftrightarrow (-{\cal L},\omega_{\rm B})$, where the Lorentz force acting on the charged particle is repulsive and has the same direction as the centrifugal force, and $({\cal L},\omega_{\rm B}) \longleftrightarrow (-{\cal L},-\omega_{\rm B})$, where the Lorentz force is attractive and has the same direction as the gravitational force. Below we analyze the effective potential (\ref{Effpot}) for positive angular momenta of the particle and either positive or negative magnetic coupling parameter.

\begin{figure}[ht!]
 \centering
    \includegraphics[width=0.9\linewidth]{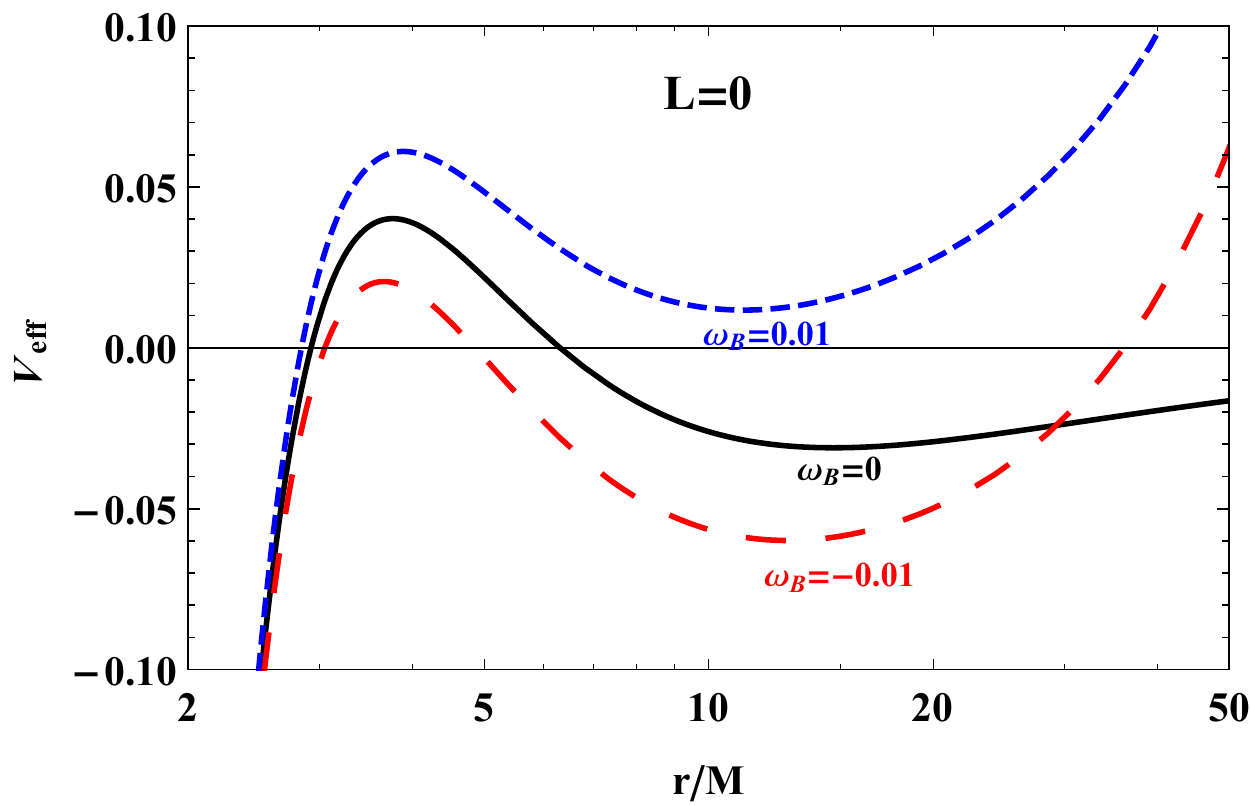}
     \includegraphics[width=0.9\linewidth]{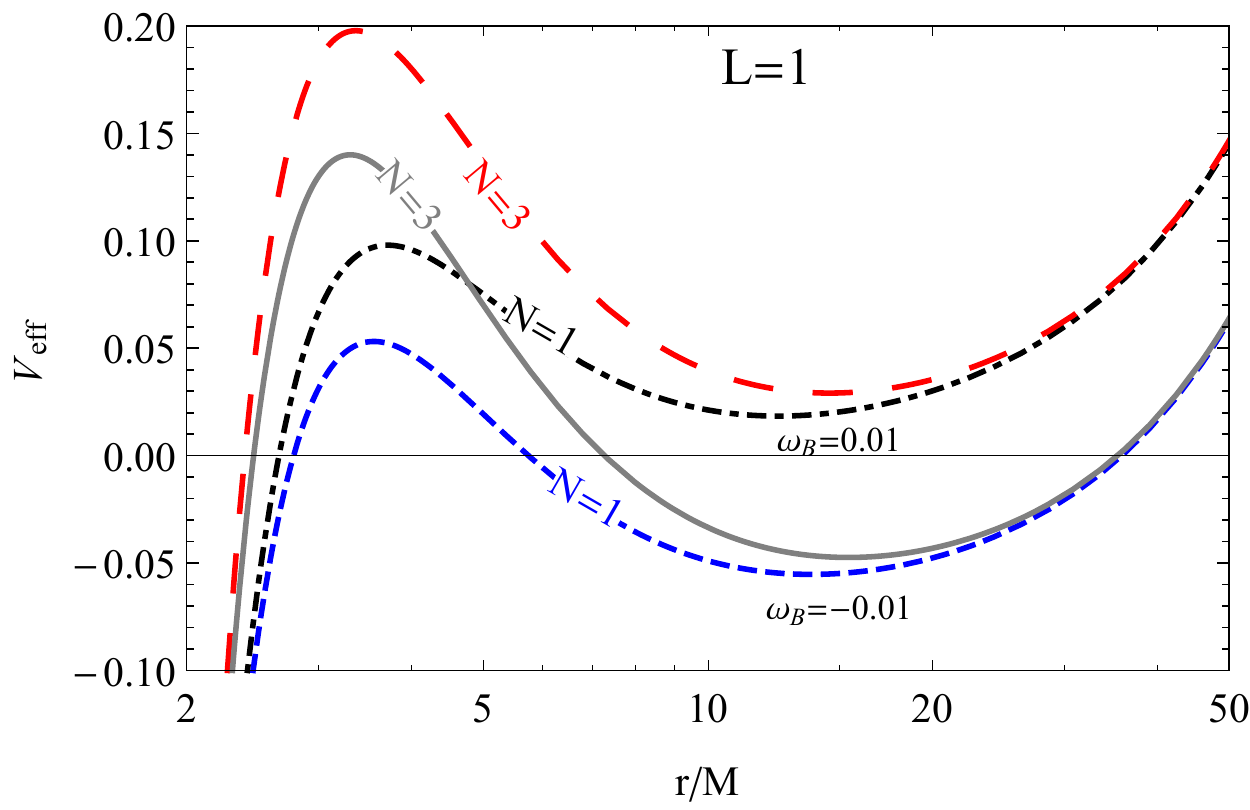}
    \caption{Radial dependence of the effective potential of a charged particle for different values of the conformal, scale, and magnetic field parameters.}
    \label{effpotfig}
\end{figure}
Fig.~\ref{effpotfig} shows the radial dependence of the effective potential on the equatorial plane. One can see from Fig.~\ref{effpotfig} that when $\omega_{\rm B}>0$ the effective potential is higher than in the case $\omega_{\rm B}<0$ and increases with the increase of the values of parameters $L$ and $N$. It is worth to note that at large distances the effect of the magnetic field plays a more important role than the effect of conformal gravity.

\subsection{Stable Circular orbits}

Now we will consider the innermost stable circular orbits of charged particles using following the standard conditions
\begin{eqnarray}\label{condition}
V_{\rm eff}'(r)=0  \ ,
\qquad
V_{\rm eff}''(r)=0 \ ,
\end{eqnarray}
In fact, circular orbits can be stable when the second derivative of the effective potential with respect to both coordinates ($\partial_{r} V_{\rm eff} \geq 0$ and $\partial_{\theta} V_{\rm eff} \geq 0$) is positive and the ISCO on the equatorial plane corresponds to the zero value of this derivative $\partial_r V_{\rm eff}=0$. The angular momentum of circular orbits can be found in the following form	

\begin{eqnarray}\label{lcriteq}
\nonumber
&&{\cal L}_{\rm cr}^{\pm}=\frac{1}{4 f(r)-2 r f'(r)} \Bigg\{r^3 \omega _B \left[S(r) f'(r)+f(r) S'(r)\right]
\\\nonumber
&&\pm r^{3/2} \Bigg[4\Big\{f(r) f'(r) \left[2 S(r)-r S'(r)\right]-r S(r) f'(r)^2\Big\}
\\
&&+f^2(r) \left\{r \omega _B^2 \left[2 S(r)+r S'(r)\right]^2+8 S'(r)\right\}  \Bigg]^{1/2} \Bigg\}\ .
\end{eqnarray}

In order to ensure that we obtain a real solution of equation (\ref{lcriteq}), we require the function under the square root to be always positive. Since the second part of the equation under the square root is always positive, it implies
\begin{eqnarray}
f(r) f'(r) \left[2 S(r)-r S'(r)\right]-r S(r) f'(r)^2 > 0\ ,
\end{eqnarray}
must be satisfied for any values of $L$ and $N$.

Now we will analyze the distance where ${\cal L}_{\rm cr}$ is always positive i.e. positions where circular motion can occurs in the equatorial plane.

  \begin{figure}[ht!]
    \centering
    \includegraphics[width=0.98\linewidth]{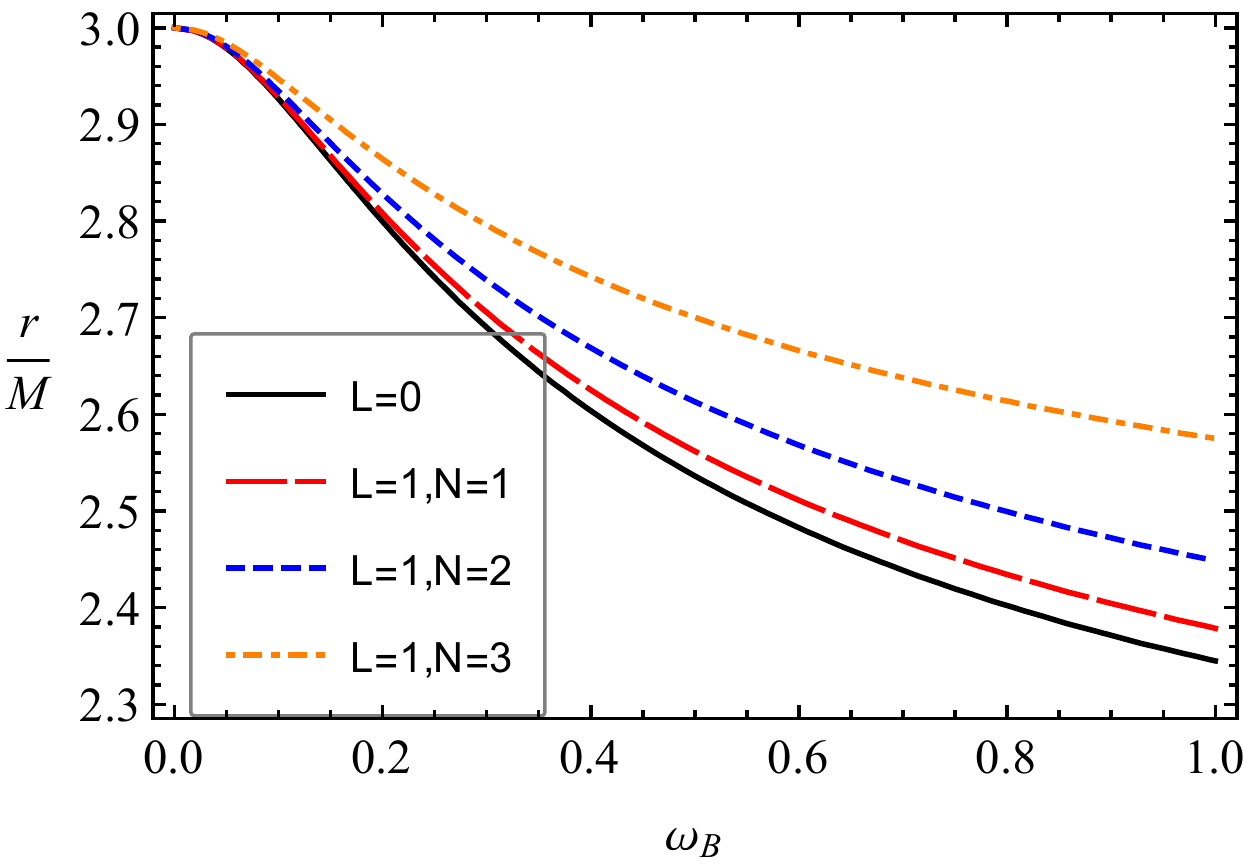}
         \caption{{Dependence of the minimum circular orbits from the magnetic interaction parameter (cyclotron frequency) $\omega_B$ for the different values of the conformal parameters $L$ and $N$ with the comparison of pure Schwarzschild case (black solid line). } }
    \label{cond3d}
    \end{figure}

{ Fig.~\ref{cond3d} illustrates the dependence of the minimum distance where circular orbits are allowed from the cyclotron frequency for the different values of the conformal parameters $L$  and $N$.  One can see from the figure that independently from the values of $L$ and $N$ such minimum orbits starts from the value $3M$ in the absence of external magnetic field and then decreases with the increase of the latter one. It is also worth to note that the rate of such decrease reduces with the increase of both conformal parameters.}

\begin{figure}[h!]
        \includegraphics[width=0.98\linewidth]{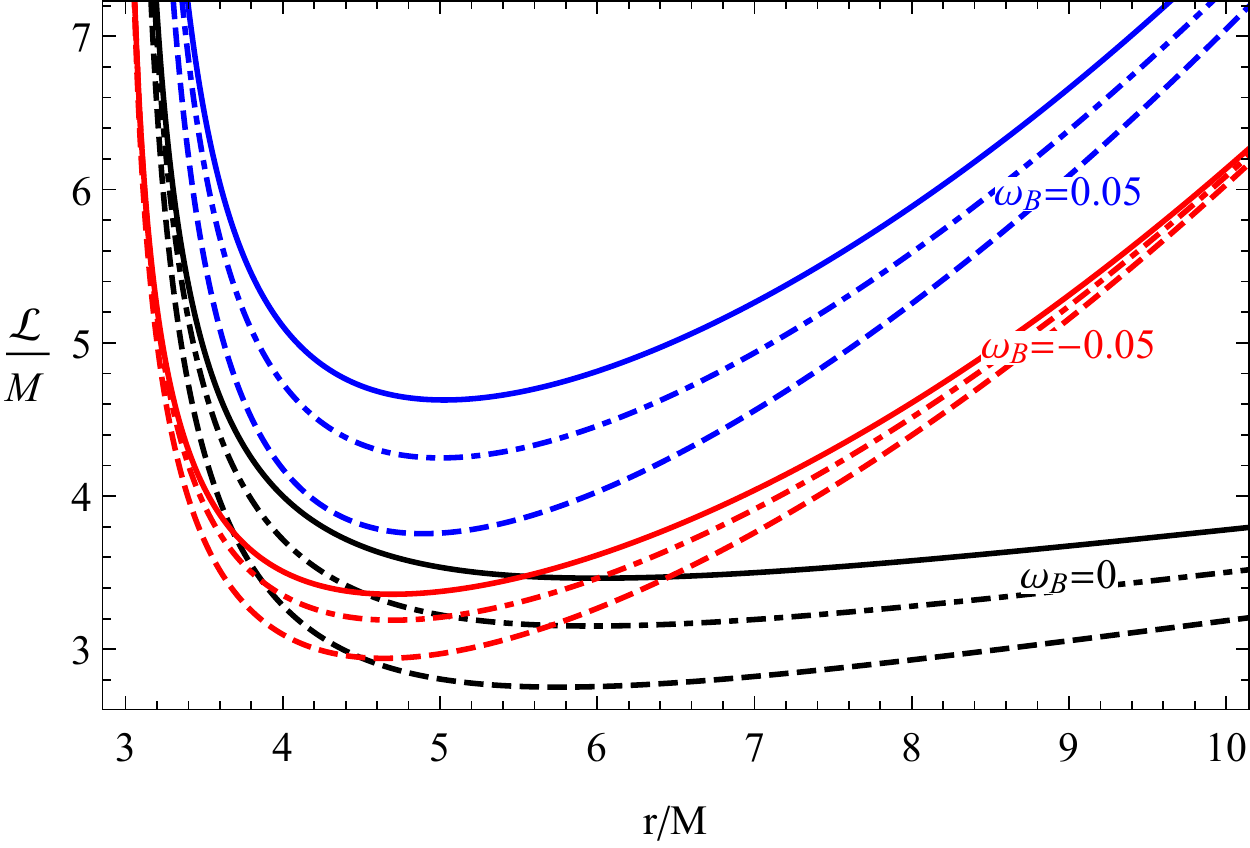}
       \caption{Radial dependence of the specific angular momentum. Solid, dashed and dot-dashed lines correspond to the values of the conformal parameters $(L,N)=(0,3)$, $(1,3)$, and $(1,1)$, respectively. Blue, red and black lines correspond to positive, negative, and vanishing charge of the particle, respectively.}
    \label{lcritfig}
\end{figure}

Fig. \ref{lcritfig} shows the radial dependence of the critical value of the angular momentum for circular orbits. One can see from this figure that the value of the critical angular momentum of the charged particle increases in the presence of an external magnetic field.

The energy of the charged particle at circular orbits can be obtained substituting Eq.~(\ref{lcriteq}) into Eq.~(\ref{Effpot})
\begin{equation}
    {\cal E}=f(r)\left[S(r)+\left(\frac{{\cal L}_{\rm cr}}{r }-S(r)\omega_{\rm B}r  \right)^2\right]\ .
\end{equation}
Here we will study radial dependence of the energy in the equatorial plane, where $\sin\theta=1$.
\begin{figure}[h!]
    \centering
    \includegraphics[width=0.99\linewidth]{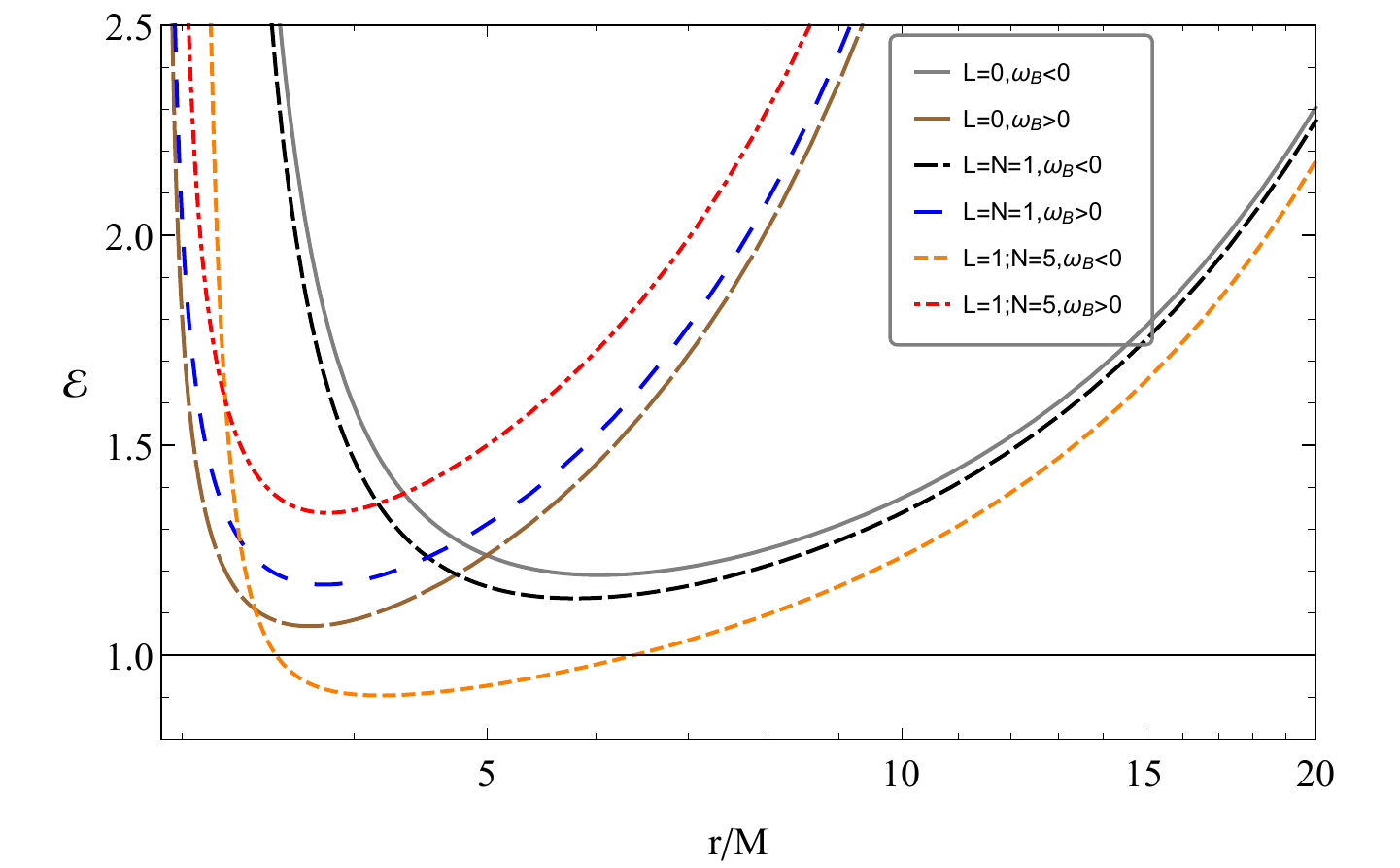}
          \caption{Radial dependence of specific energy of the charged particle with magnetic interaction parameter $|\omega_B|=0.1$ at circular orbits }
    \label{veffco}
\end{figure}

Fig.~\ref{veffco} illustrates the radial profiles of the charged particle energy in circular orbits on the equatorial plane. One can see that the magnetic field increases the energy of the charged particle. One more thing is that the rate of energy increase in the case of $\omega_B>0$ is higher than the case of $\omega_B<0$.


\begin{figure}[h!]
  \centering
    \includegraphics[width=0.98\linewidth]{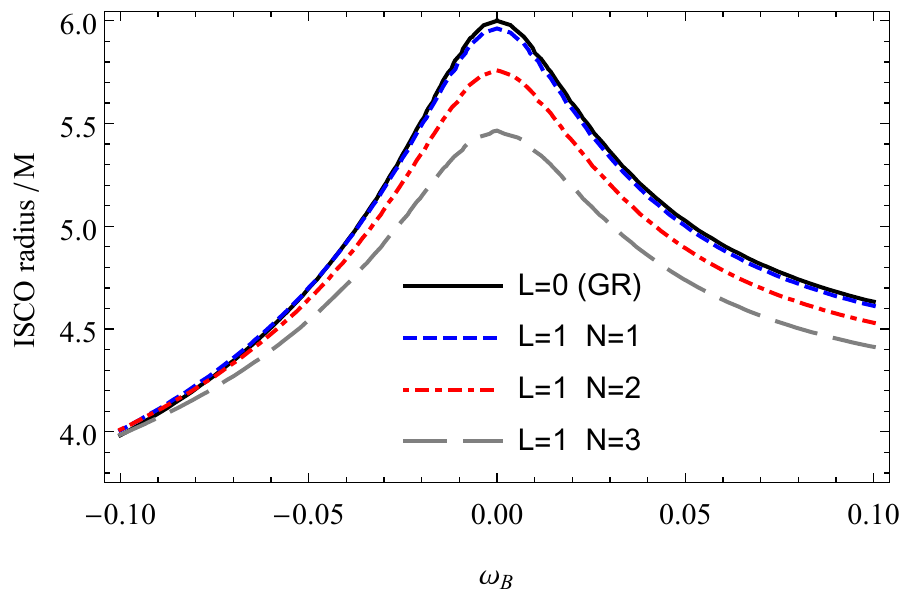}
           \caption{Dependence of the ISCO radius of the charged particle on the magnetic parameter for various values of the conformal and scale parameters.}
    \label{iscoo}
\end{figure}

In Fig.~\ref{iscoo}, we present the effect of conformal gravity and magnetic field on the ISCO radius of charged particles. One can see from the figure that increasing the conformal parameters $L$ and $N$ and the magnetic coupling parameter decreases of the ISCO radius.


\begin{figure}[h!]
  \centering
    \includegraphics[width=0.98\linewidth]{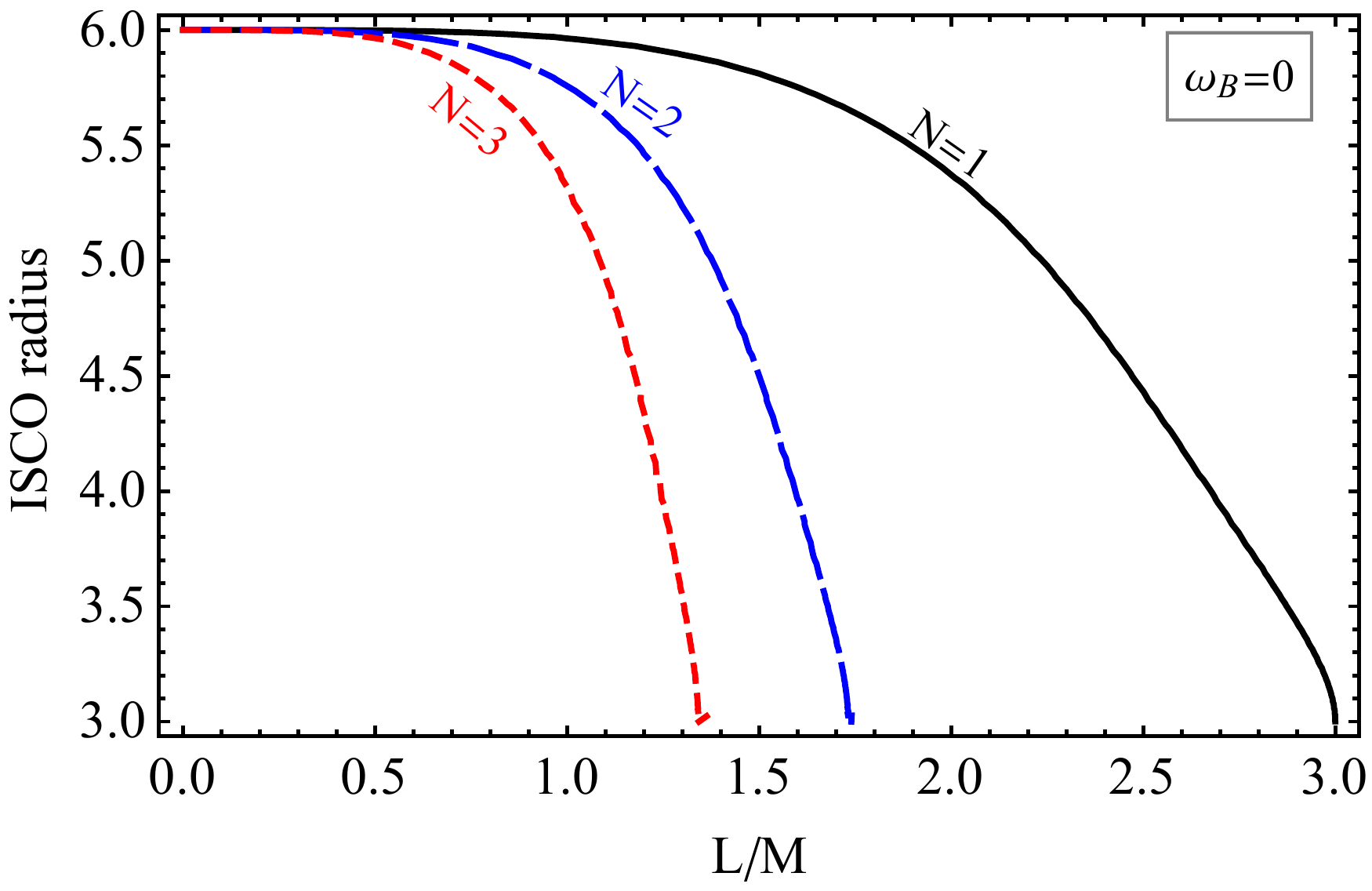}
    \includegraphics[width=0.98\linewidth]{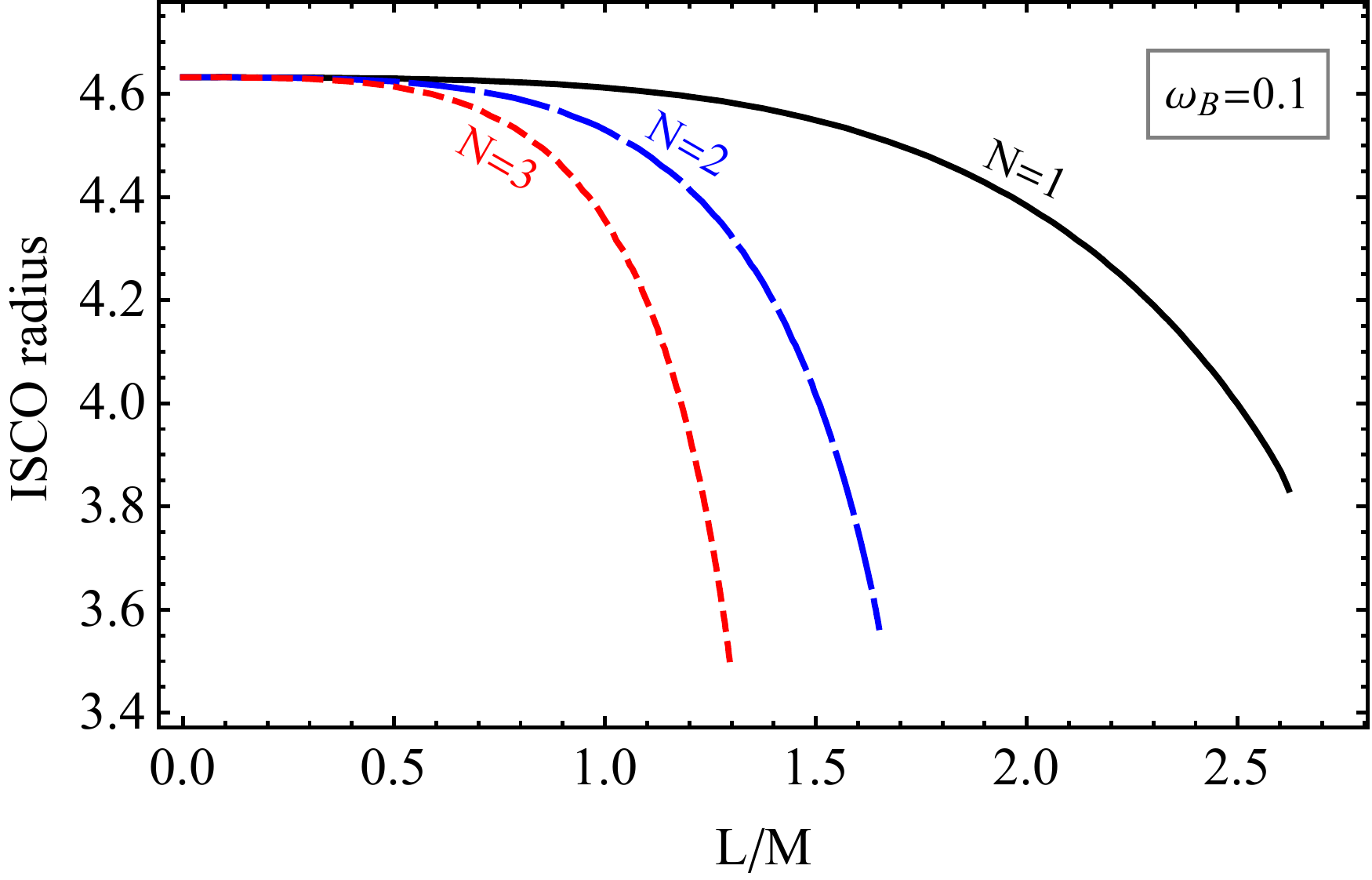}
           \caption{{Dependence of the ISCO radius of neutral and charged particles from the  conformal parameter $L$ for various values of the parameter $N$.}}
    \label{iscoL}
\end{figure}

{In Fig.~\ref{iscoL} it is illustrated the dependence of ISCO radius of neutral (first panel) and charged (second panel) particles on the conformal parameter $L$, for the chosen values of the parameter $N$. One can see from the first panel that the minimum values of the ISCO radius tends to $3M$ for the neutral particles independently from the parameter $N$, while for the charged ones this minimum ISCO radius decreases with the increase of $N$.}  

\subsection{Charged particles trajectories}

Now we will study the effects of the parameters of conformal gravity on the trajectories of charged particles. 

\begin{figure*}[ht!]
\centering
\includegraphics[width=0.99\linewidth]{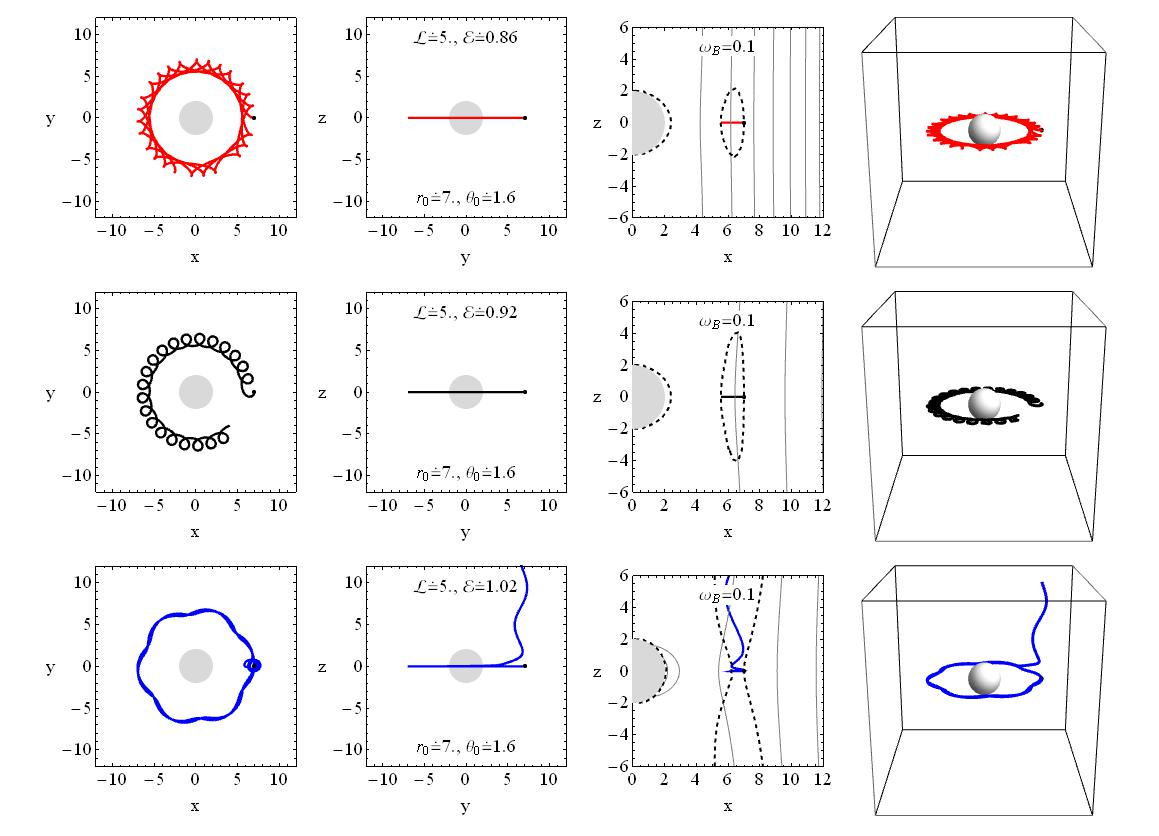}
\caption{Trajectories of positively charged particles around a non-rotating black hole immersed in a magnetic field in conformal gravity. Red, black and blue trajectories correspond to the case $N=1$, 4, and 8, respectively, when the conformal parameter $L/M=1$ started from equatorial plane and the distance $r_0=7M$. All trajectories have been plotted for the value of specific angular momentum ${\cal L}=5$. Black dashed lines and gray solid lines correspond to magnetic field lines. }
\label{trpos}
\end{figure*}

\begin{figure*}[ht!]
\centering
\includegraphics[width=0.99\linewidth]{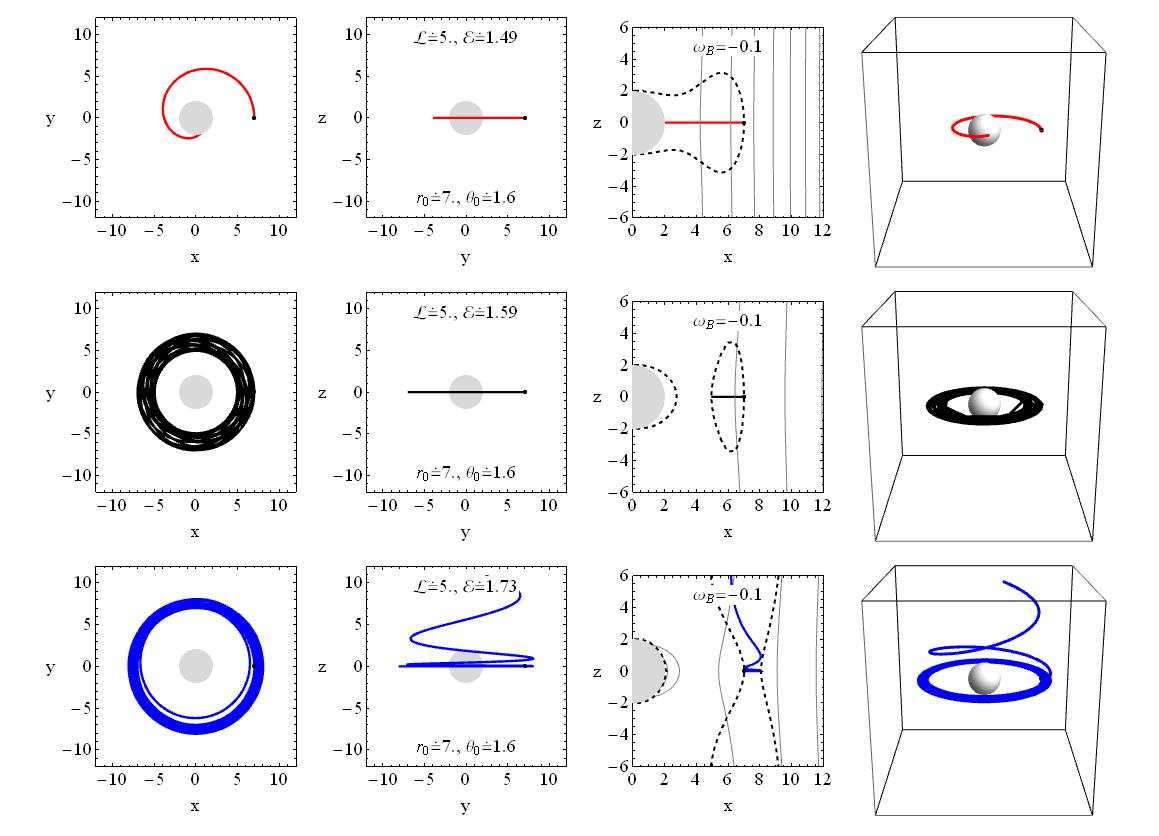}
\caption{As in Fig.\ref{trpos}, but for negatively charged particles. }
\label{trneg}
\end{figure*}

Trajectories of positive and negative charged particles for absolute values of the magnetic coupling parameter $|\omega_B|=0.1$ are shown in Fig. \ref{trpos}. All plots in this figure are taken at the value of the conformal parameter $L=M$ and trajectories of the charged particles in red lines on first and fourth rows correspond to the values of the conformal parameters $N=1$, black and blue lines correspond to the conformal parameter $N=4$ and $N=8$, respectively for the fixed value of the specific angular momentum ${\cal L}=5M$. One can see from the figures that the specific energy of the charged particles increases with increasing of the conformal parameter $N$. Moreover, the orbits of charged particles started at equatorial plane $\theta_0=\pi/2$ become unstable and the particle leaves the central object at higher values of $N$. It can be explained by the magnetic field structure around the black hole in conformal gravity which is shown in the third column of the plots in grey lines that becomes dipol like structure at the higher values of the parameter $N$. 


\subsection{Conformal non-rotating black hole versus Schwarzschild black hole in a uniform magnetic field}

In this subsection, we consider two different cases: the motion of a charged particle around a Schwarzschild black hole and a non-rotating black hole in conformal gravity, in a magnetic field, with the same ISCO radius for the charged particle.

\begin{figure}[ht!]
   \centering
    \includegraphics[width=0.98\linewidth]{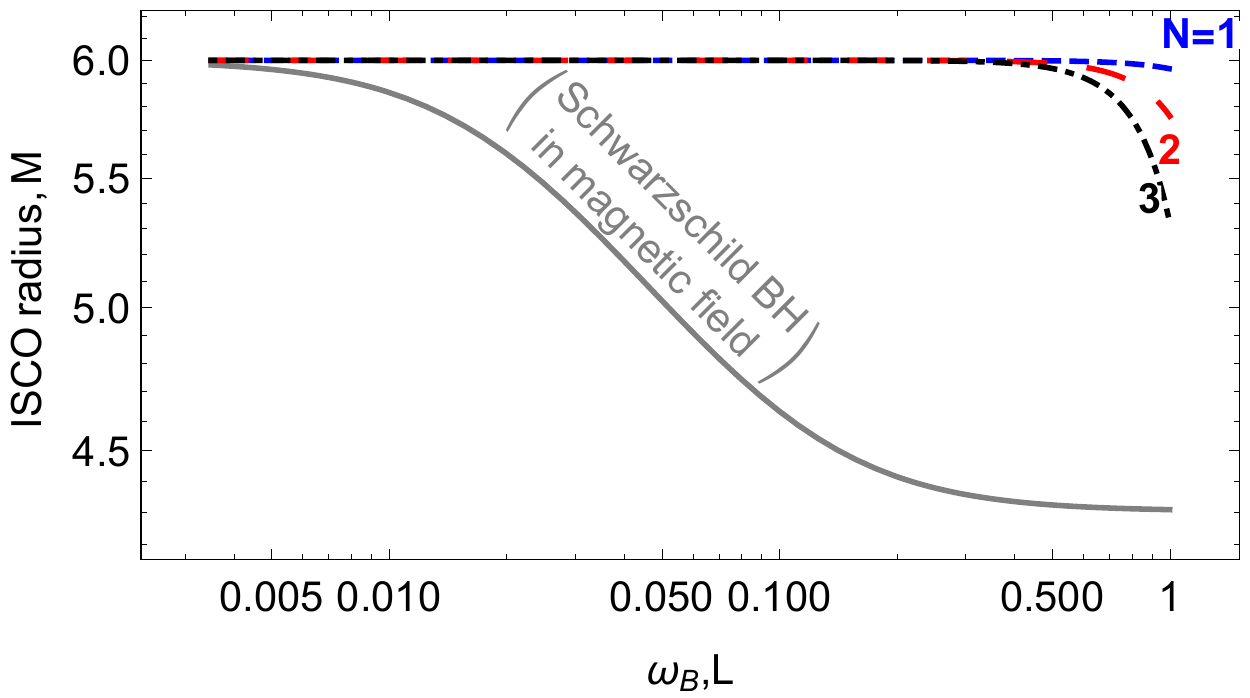}
           \caption{Dependence of the ISCO radius of charged particle on the magnetic coupling parameter $\omega_{\rm B}$ and conformal parameter $L$ for different values of the parameter $N$. Units in which $M=1$.}
    \label{iscoLw}
\end{figure}

Fig.~\ref{iscoLw} show that the impact of the conformal parameter $L$ and of the magnetic coupling parameter $\omega_{\rm B}$ on the ISCO radius is the same.

\begin{figure}[ht!]
   \centering
    \includegraphics[width=0.98\linewidth]{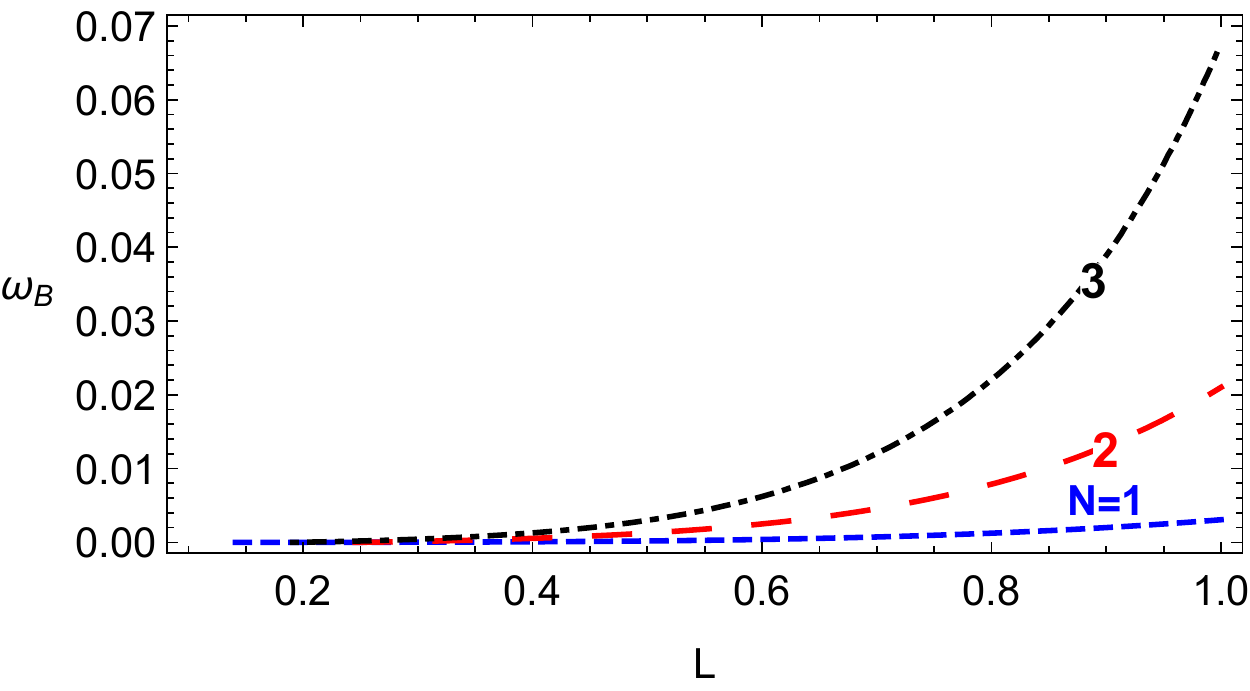}
           \caption{Relation between the conformal parameter $L$ and the magnetic coupling parameter $\omega_{\rm B}$ for the same ISCO radius of charged particles and different values of the conformal parameter $N$.}
    \label{Lw}
\end{figure}

Fig.~\ref{Lw} shows the relation between the conformal and magnetic coupling parameters $L$ and $\omega_B$, for the same ISCO radius. One can see from this figure that a magnetic field can mimic conformal gravity for the values of the parameter $L=1$ and the other conformal parameter $N=1,2,3$  at $\omega_B \leq 0.003068$, $\omega_B\leq 0.021015$ and $\omega_B\leq 0.06771$, respectively.

\section{\label{sec_3} Charged particle motion around Conformal rotating black holes immersed in a uniform magnetic field. }

Since the spacetime of a rotating black hole in conformal gravity
admits separation of variables on the equatorial plane ($\theta=\pi/2$) we will study the motion around the source described by the metric (\ref{metric1}) using the Hamilton-Jacobi equation where the
action ${\cal S}$ can be decomposed in the form as Eq.~(\ref{action}). Finally we obtain the equation of motion in the following form (for $N=1$):
\begin{eqnarray}
\dot{t}&=& e B a
+\frac{  {\cal E} r^3 -2 a {\cal L} M  + a^2 {\cal E} (2 M  + r)}{r^{-7}\Delta (L^2 + r^2)^4 } \\
\dot{r}^2&=&R(r) \\
\dot{\phi}&=& \frac{e B}{2}+\frac{ 2 a {\cal E} M  + {\cal L} (r - 2 M )}{r^{-7}\Delta(L^2 + r^2)^4 }
\end{eqnarray}
where 
\begin{widetext}
\begin{eqnarray}\label{vr}
R(r)&=&\frac{\Delta }{ \Sigma S(r)} \left[\left(\frac{\omega_{\rm B} S(r)
   \left(\left(a^2+r^2\right) \Sigma -2 a^2 M r\right)}{ \Sigma}-{\cal L}\right) \left(\frac{(2 M r (a {\cal E} -{\cal L})+{\cal L} \Sigma )}{S(r) \left(\Sigma 
   \left(a^2+r^2\right)-2 M r^3\right)}-\omega_{\rm B}\right)-1 \right.
\\\nonumber   
 &&  \left.+\frac{ \left(\frac{2 M r \Sigma ^4
   \left(a^2 {\cal E} -a {\cal L}+{\cal E}  r^2\right)}{2 M r^3-\Sigma  \left(a^2+r^2\right)}+2 a \omega_{\rm B}
   L^2 (L^6+4 L^4 \Sigma +6 L^2 \Sigma ^2+4  \Sigma ^3) + (2 a \omega_{\rm B} -{\cal E})  \Sigma ^4\right)}{S(r) \Sigma^5 \left(2 a \omega_{\rm B} S(r) (\Sigma -M r)-{\cal E} \Sigma \right)^{-1}}\right]
\end{eqnarray}
\end{widetext}

One may define the effective potential in the following form
\begin{equation}
V_{eff}(r) = \frac{{\cal E}^2-1-R(r)}{2}\ .
\end{equation}

\begin{figure*}[ht!]
\begin{center}
a.
\includegraphics[width=0.31\linewidth]{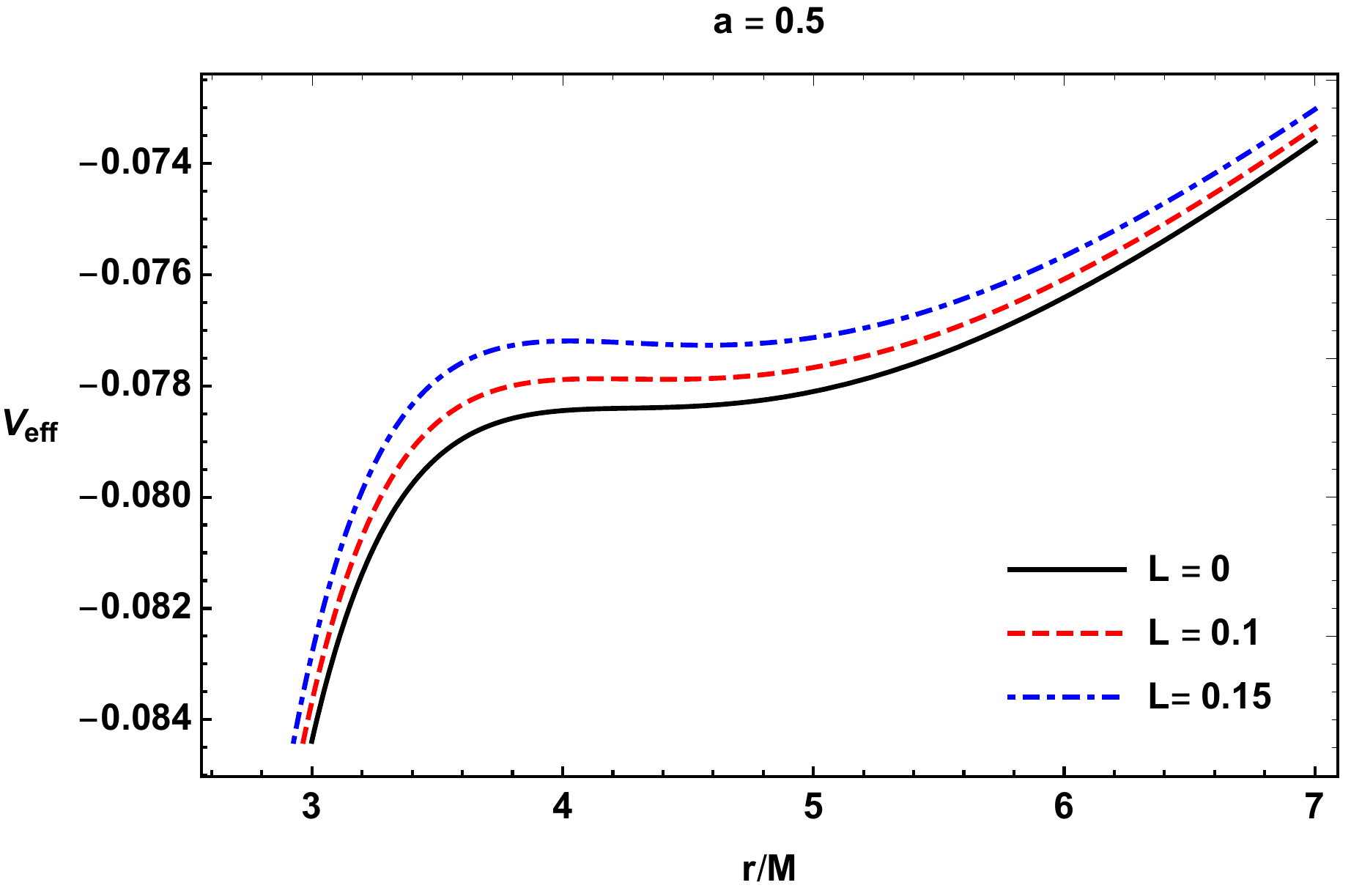}
b.
\includegraphics[width=0.31\linewidth]{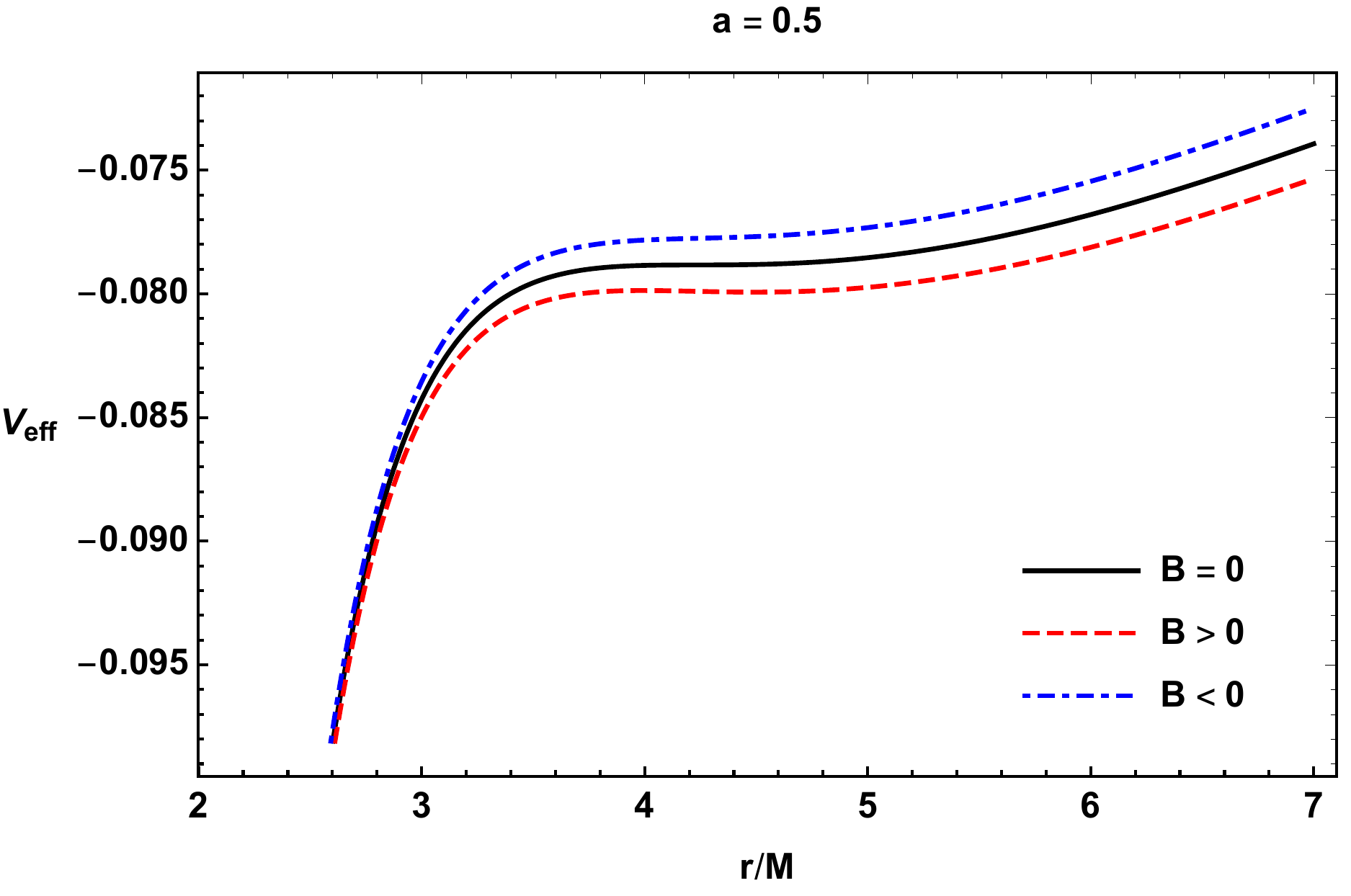}
c.
\includegraphics[width=0.3\linewidth]{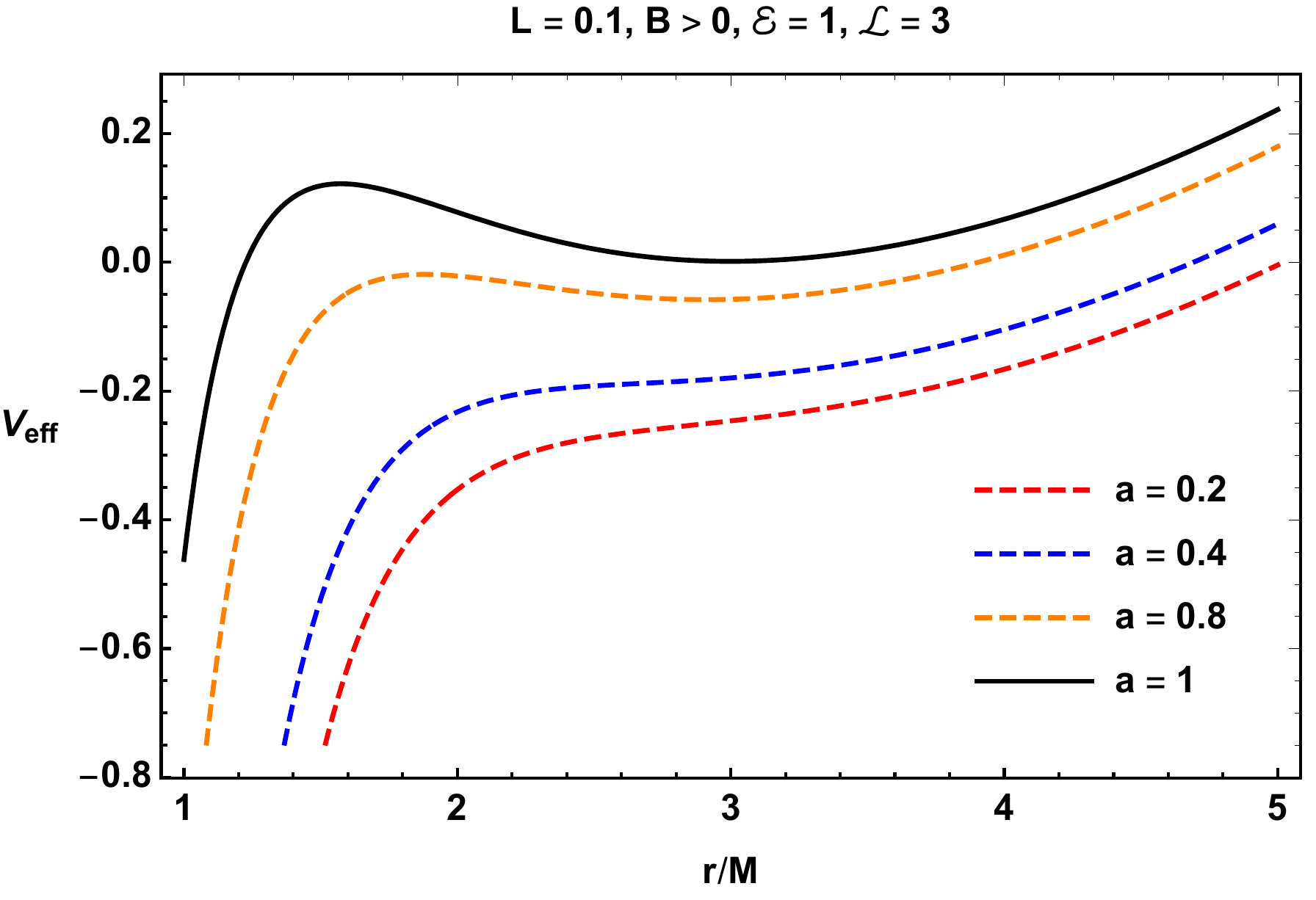}
\end{center}
\caption{Radial dependence of the effective potential of
radial motion of a charged particle around rotating quasi-Kerr black
hole in the equatorial plane. The figures correspond to the case of
a. without external magnetic field, b. in the presence of external magnetic field when $L=0.05$ and c. in the presence of both parameters. Units in which $M=1$. \label{fig2}}
\end{figure*}

{ Radial dependence of the effective potential is presented in Fig.~\ref{fig2}, where graphs a. and b. are plotted in the case when the values of the angular momentum ${\cal L}$ and energy ${\cal E}$ of the particle are chosen to be equal to the values of a particle moving around the innermost circular orbit so, turning points of the lines on the graphs represent the ISCO radius; while in c. the lines correspond to the case when energy and angular momentum of the test particle are fixed for the various values of rotation parameter of a black hole. }

\begin{figure*}[ht!]
\begin{center}
a.
\includegraphics[width=0.30\linewidth]{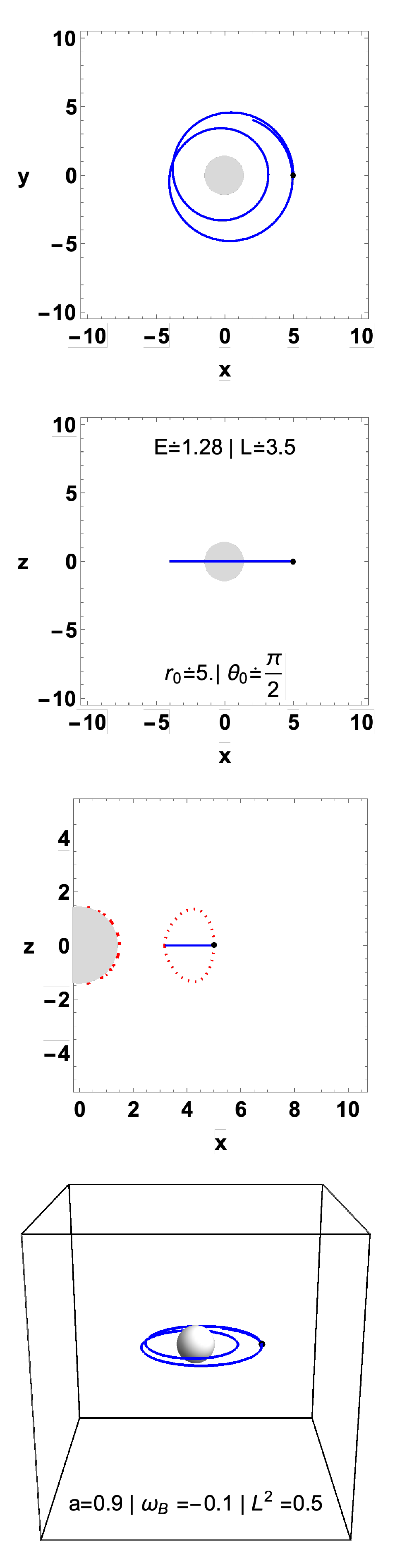}
b.
\includegraphics[width=0.30\linewidth]{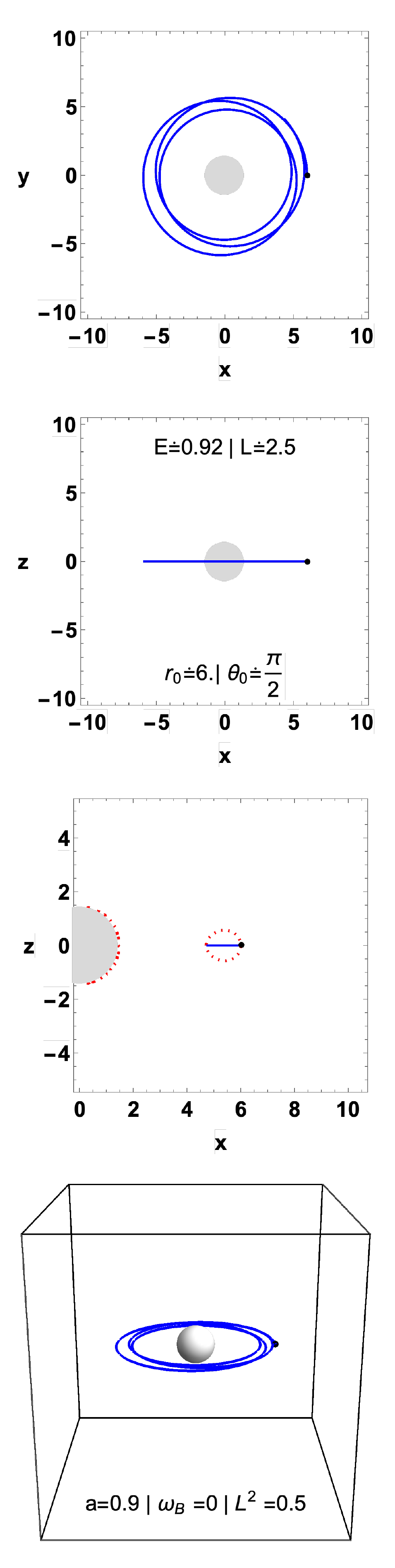}
c.
\includegraphics[width=0.30\linewidth]{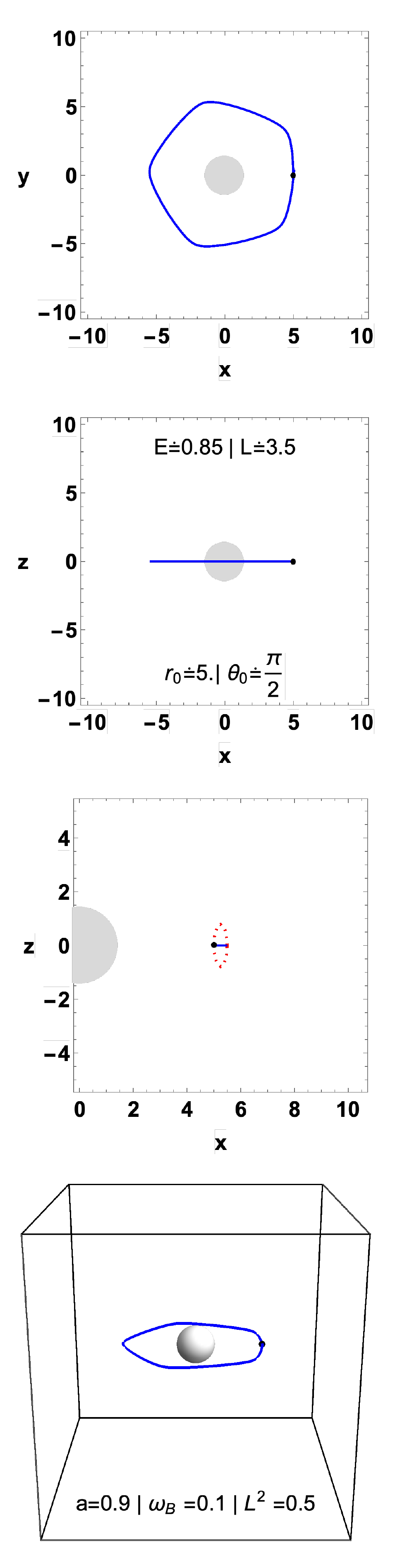}
\end{center}
\caption{Trajectories of charged particles on the equatorial plane $\theta_{0}=\pi/2$ for given values of rotation, magnetic and conformal parameters (the values of the parameters are the same for each column and written in the last row). Units in which $M=1$.\label{trj}}
\end{figure*}

The stability of the equatorial orbits can be checked by plotting the trajectories of charged particles for given values of the external magnetic field and the conformal parameter $L$ as illustrated in Fig.~\ref{trj} (the $z$ axis is assumed to be parallel to the symmetry axis and the origin coincides with the centre of the gravitating object). It is clearly seen from the second row of figures that for fixed values of the parameters mentioned the trajectory remains stable. 

\begin{figure*}[ht!]
\begin{center}
a.	
\includegraphics[width=0.47\linewidth]{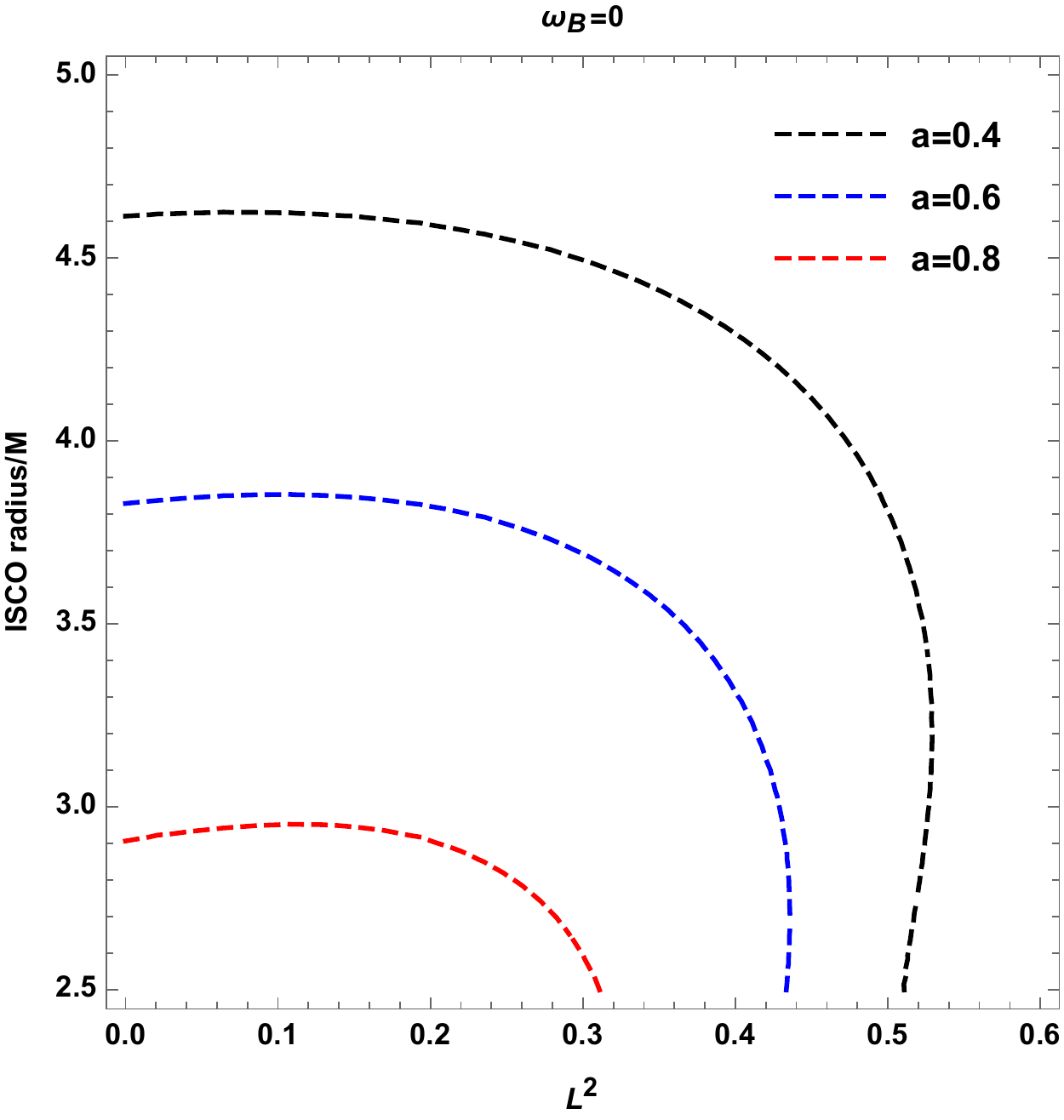}
b.
\includegraphics[width=0.47\linewidth]{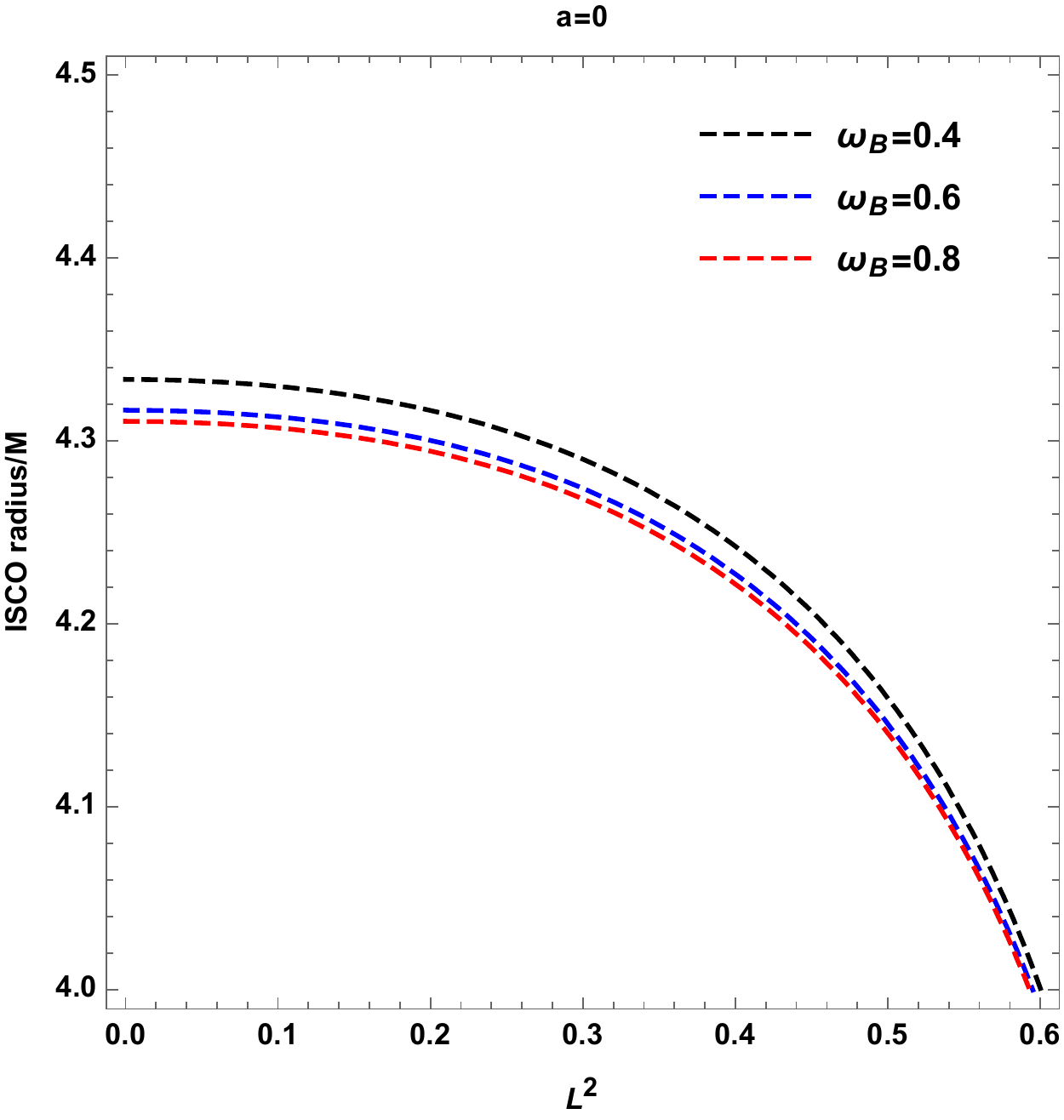}
\end{center}
	\caption{Dependence of the ISCO radius on the conformal gravity parameter $L^2$: a. in the absence of an external magnetic field and b. in the absence of rotation. Units in which $M=1$. \label{isco_1}}
\end{figure*}

The value of the ISCO radius can be obtained from the following standard conditions
\begin{eqnarray}
 \label{min}
 R(r)&=&0 \ ,
\\\label{circle}
R'(r)&=&0 \ , 
\\\label{isco}
R''(r)&=&0 \ , 
\end{eqnarray}
with $R(r)$ taking the form (\ref{vr}).
The results of equations (\ref{min})-(\ref{isco}) are expressed  in Fig.~\ref{isco_1}, panel~a for a vanishing magnetic field and a non-vanishing spin parameter and panel~b for a non-vanishing magnetic field and a vanishing spin parameter. We can clearly see from the figures that in the absence of an external magnetic field but non-vanishing spin parameter the ISCO radius first slightly increases for small values of the conformal parameter and then goes down for higher values of $L^2$. In the case of a vanishing spin parameter but in the presence of an external magnetic field, it always monotonically decreases with the increase of the conformal parameter $L^2$. In Tab.~\ref{1tab} and Tab.~\ref{2tab} numerical results in the presence of both rotation and magnetic parameters are presented. From the tables, one can see that for chosen values of $\omega_B$ the ISCO radius increases if one increases the conformal gravity parameter for a given interval of the latter one. The opposite scenario takes place for a given value of conformal parameter i.e. the increase of magnetic parameter decreases the ISCO radius. { One should also mention that for the case of extreme rotation the ISCO radius becomes bigger compared to the case of pure Kerr metric where the ISCO becomes close to $M$.}

\begin{table}
\caption{\label{1tab} The innermost stable circular orbits of particles moving around a rotating black hole
(case for $a=0.5$ and $M=1$). }
\begin{ruledtabular}
\begin{tabular}{|c|c|c|c|c|c|}
 $L^2$ & 0 & 0.001 & 0.002 & 0.005 & 0.01 \\
\hline
$\omega_{\rm B}=0.05$ & 3.7312 & 3.7314 & 3.7316 & 3.7322 & 3.7331
\\
\hline
$\omega_{\rm B}=0.1$ & 3.2897 & 3.2899 & 3.29 & 3.2904 & 3.2911
\\
\hline
$\omega_{\rm B}=0.2$ & 2.8444 & 2.8445 & 2.8446 & 2.8449 & 2.8454
\\
\hline
$\omega_{\rm B}=0.4$ & 2.4799 & 2.48 & 2.4801 & 2.4803 & 2.4807
\\
\hline
$\omega_{\rm B}=0.8$ & 2.2192 & 2.2193 & 2.2194 & 2.2195 & 2.2197
\\
\end{tabular}
\end{ruledtabular}
\end{table}

\begin{table}
\caption{\label{2tab} The innermost stable circular orbits of particles moving around a rotating black hole
(case for $\omega_B=0.25$ and $M=1$). }
\begin{ruledtabular}
\begin{tabular}{|c|c|c|c|c|c|}
 $L^2$ & 0 & 0.001 & 0.002 & 0.005 & 0.01 \\
\hline
$a=0.05$ & 3.1458 & 3.1459 & 3.146 & 3.1462 & 3.1467
\\
\hline
$a=0.1$ & 3.1074 & 3.1075 & 3.1076 & 3.1078 & 3.1083
\\
\hline
$a=0.2$ & 3.0248 & 3.0249 & 3.025 & 3.0253 & 3.0257
\\
\hline
$a=0.4$ & 2.8316 & 2.83165 & 2.8317 & 2.832 & 2.8324
\\
\hline
$a=0.8$ & 2.2265 & 2.2267 & 2.2268 & 2.2273 & 2.2281
\\
\hline
\bf $a=0.99$ & 1.4045 & 1.4049 & 1.4054 & 1.4067 & 1.4092
\\
\end{tabular}
\end{ruledtabular}
\end{table}

\section{\label{Sec:collision} Center-of-mass energy of charged particles collisions }

In this section, we investigate the center-of-mass energy from collisions of two charged particles near rotating magnetized black holes in conformal gravity. The general expression for the center-of-mass energy for two particles coming from infinity with masses $m_1$ and $m_2$ and four-velocities $u_1^{\alpha}$ and $u_2^{\beta}$, respectively, can be found as the sum of their four-momenta~\cite{Grib11, Grib11a}
\begin{eqnarray}\label{cmen1}
\{E_{cm},0,0,0\}=m_1u_1^{\mu}+m_2u_2^{\mu}\ ,
\end{eqnarray}
Square of the center-of-mass energy can be defined in (\ref{cmen1}) and we have
\begin{eqnarray}
E^2_{cm}=m_1^2+m_2^2-2m_1m_2g_{\mu\nu} u_1^{\mu}u_2^{\nu}\ ,
\end{eqnarray}
after algebraic substitutions, we have the expression in a dimensionless form
\begin{equation}\label{ecm2}
\frac{E^2_{cm}}{m_1m_2}=\frac{m_1}{m_2}+\frac{m_2}{m_1}-2g_{\mu\nu} u^{\mu}_1u^{\nu}_2\ .
\end{equation}

Using the expression for the four velocities of the charged particles around magnetized rotating black holes in conformal gravity and considering the collision of the particles with equal mass $m_1=m_2$ and initial energy ${\cal E}_1={\cal E}_2=1$, one may get the expression for the center-of-mass energy in the following form
\begin{equation}\label{ecm2}
\frac{E^2_{cm}}{2m^2}={\cal E}_{cm}^2=1-g_{\mu\nu} u^{\mu}_1u^{\nu}_2\ .
\end{equation}

We analyze the center-of-mass energy of the collision of positive and positive (positive and negative) charged particles with the magnetic interaction parameter $|\omega_{\rm}|=0.1$ coming from infinity with initial energy ${\cal E}_1={\cal E}_2=1$ and angular momentum ${\cal L}_1=-{\cal L}=4$.

\begin{figure*}[ht!]
    \centering
    \includegraphics[width=0.49\linewidth]{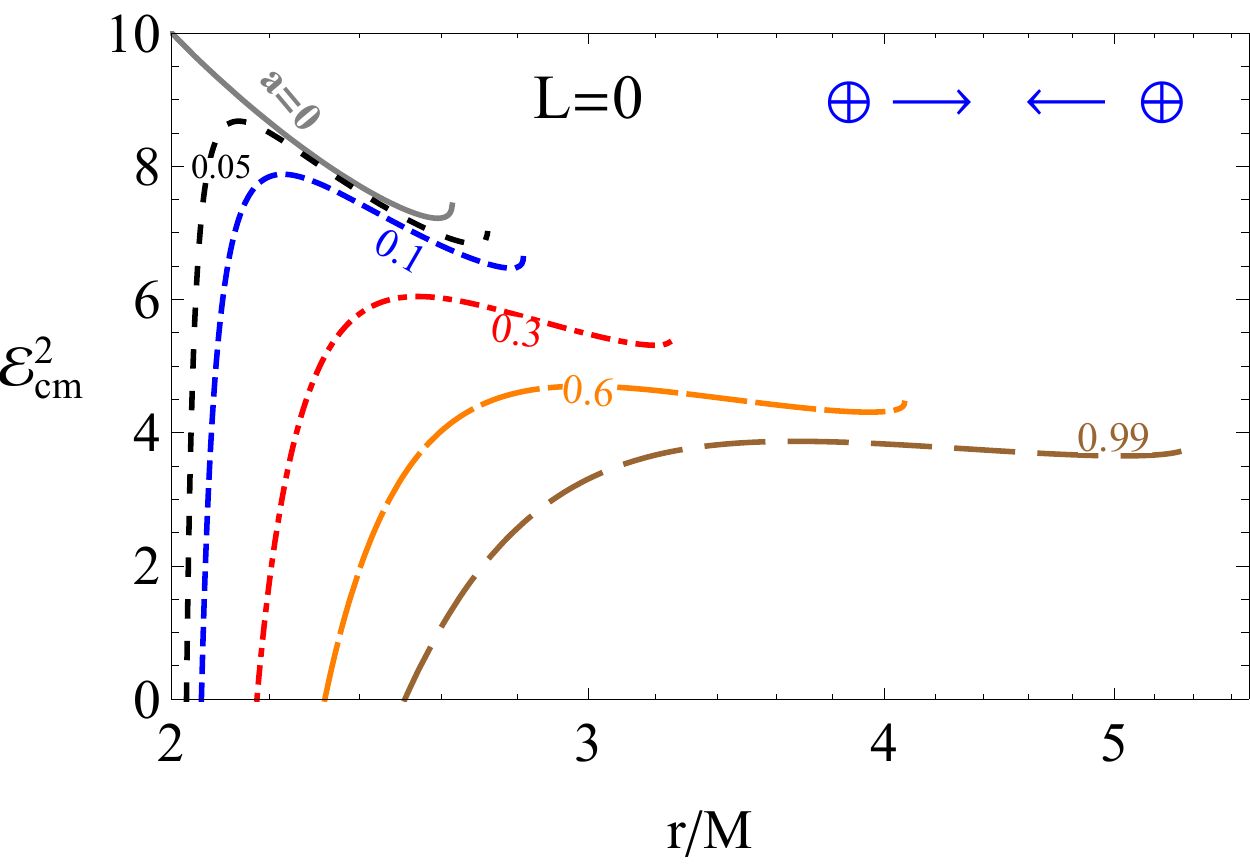}
    \includegraphics[width=0.49\linewidth]{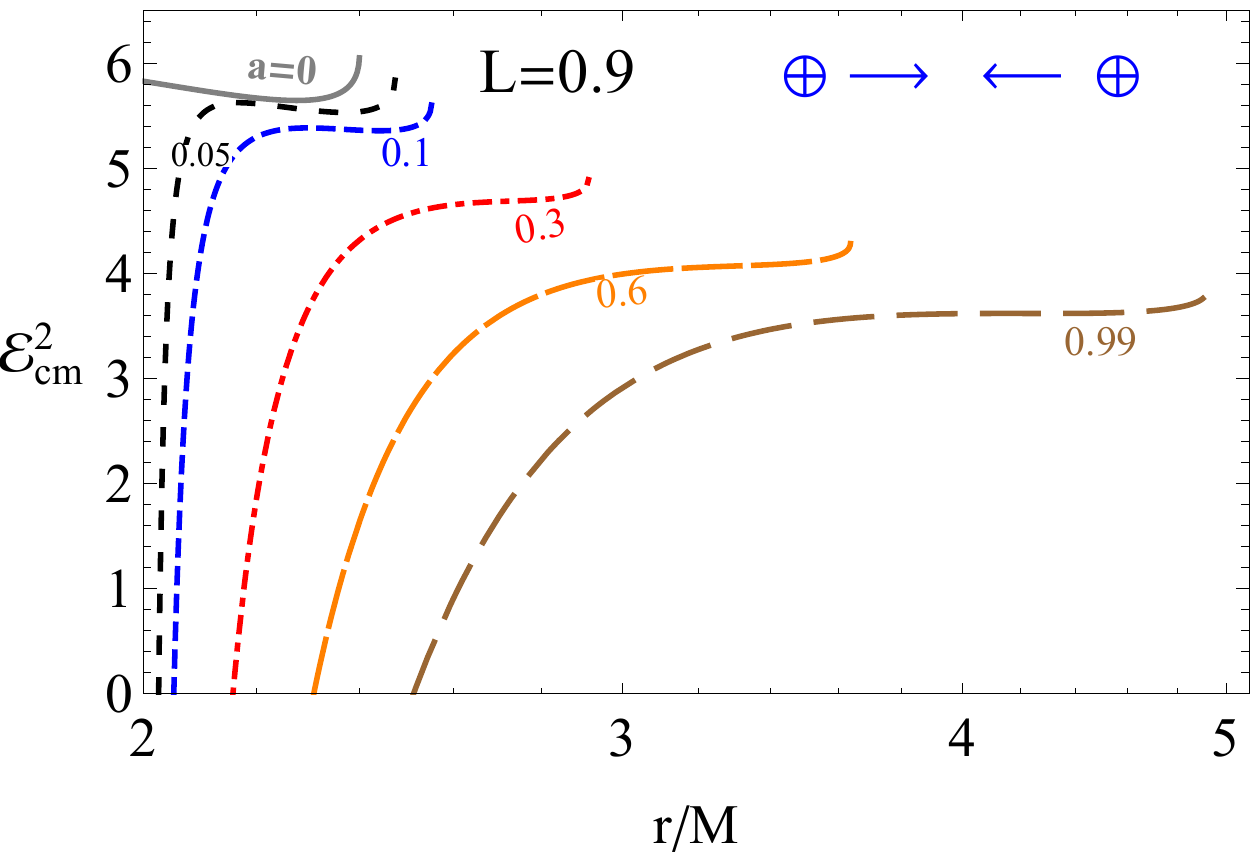}
    \includegraphics[width=0.5\linewidth]{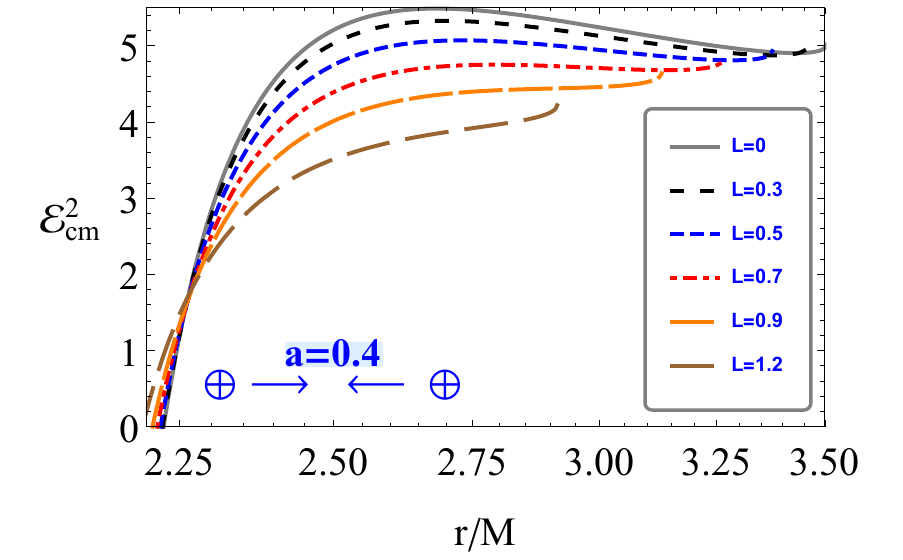}
           \caption{Radial dependence of the center-of-mass energy of two collisions of positively charged particles near a singularity-free black hole in conformal gravity for different values of the conformal parameter $L$ and the spin parameter $a$. Units in which $M=1$.}
    \label{cmenery1}
\end{figure*}

\begin{figure*}[ht!]
    \centering
    \includegraphics[width=0.49\linewidth]{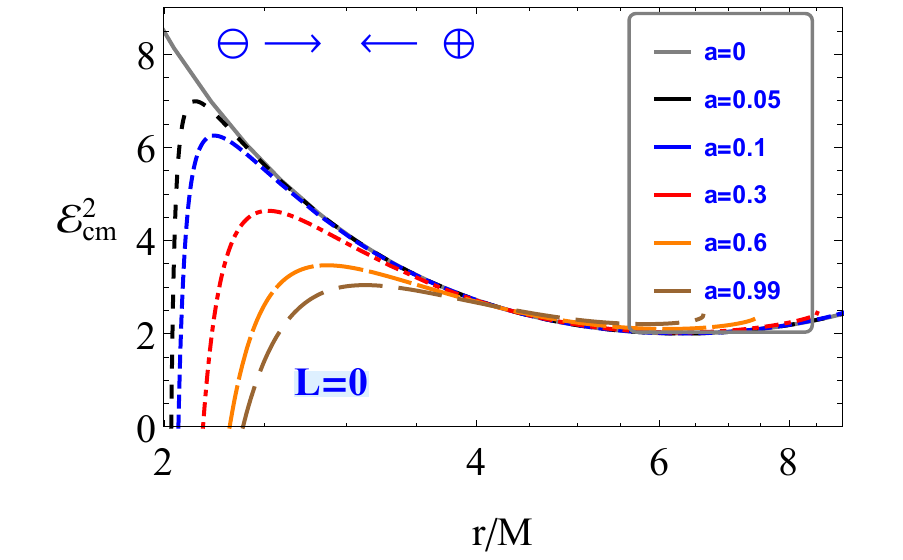}
    \includegraphics[width=0.49\linewidth]{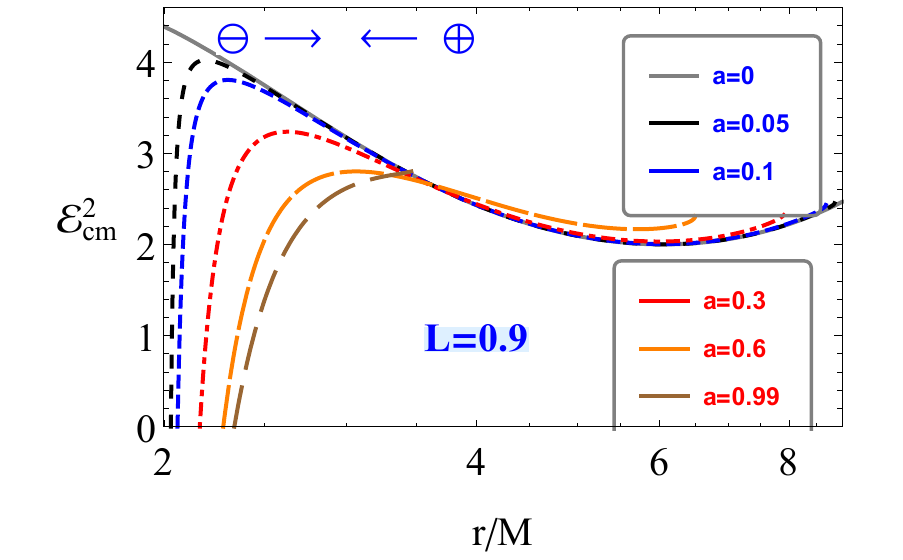}
    \includegraphics[width=0.5\linewidth]{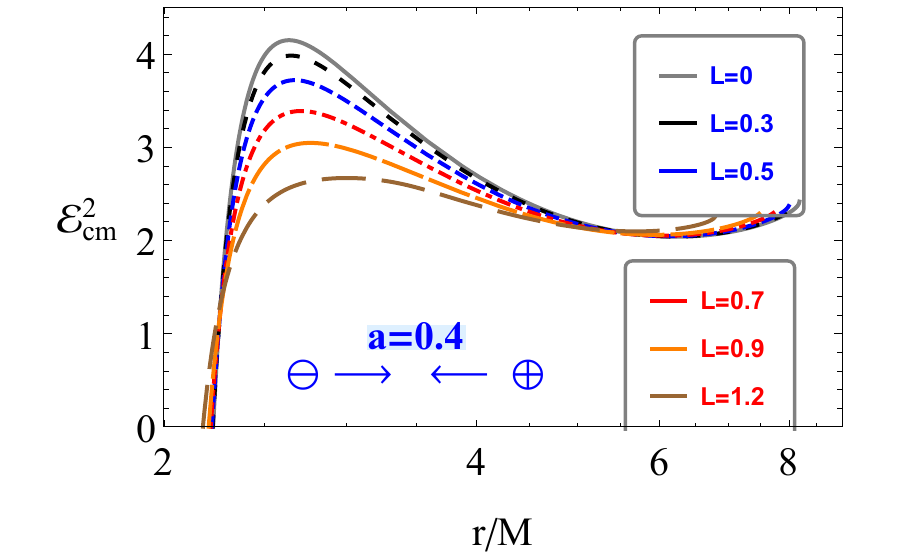}
           \caption{As in Fig.\ref{cmenery1}, but for positive and negative charged particle collisions. Units in which $M=1$.}
    \label{cmenery2}
\end{figure*}

The radial dependence of center-of-mass energy of collisions of positive-positive and positive-negative charged particles near a rotating black hole in an external magnetic field is shown in Figs.~\ref{cmenery1} and \ref{cmenery2}, for different values of the conformal and spin parameters. One can see from the figures that the center of mass energy decreases with the increase of the conformal parameter $L$ and spin parameter. Moreover, in the case of the collisions of positive-positive charged particles, the center of mass energy disappears due to dominated repulsive Coulomb forces. This implies that the collision does not occur at the point where the energy disappears. The distance where the center-of-mass energy disappears increases (decreases) increasing the conformal (spin) parameter.

\section{Conclusion}
\label{Sec:Conclusion}

In this work we have considered dynamics of charged particles and electromagnetic fields in the vicinity of rotating black holes in conformal gravity immersed in an external, asymptotically uniform magnetic field. The study of electromagnetic fields shows that the angular (radial) component of the magnetic field and the absolute value of the external magnetic field decrease (increases) with the increase of both parameters of conformal gravity, $L$ and $N$, and the increase of the parameters of conformal gravity forces the external uniform magnetic field to have a dipole-like structure.

One may see from the studies of particle dynamics of charged particles around conformal non-rotating black holes in the presence of a magnetic field that the minimum circular orbits and ISCO radius decrease as with the increase of the conformal gravity and magnetic coupling parameters and in the case of rotating black holes the ISCO decreases faster. Moreover, it is shown that the particle orbits become unstable at higher values of both conformal parameter as a result of the fact that the magnetic field gets a dipole structure.

We have studied the effect of conformal gravity on the ISCO radius of charged particles around non-rotating black holes in the presence of an external magnetic field. We  have shown that the conformal parameters can mimic the magnetic coupling parameters when $\omega_{\rm B} \leq 0.003068$ ($\omega_{\rm B} \leq 0.021015$, $\omega_{\rm B} \leq 0.06771$ ) at the values of conformal parameters $N=1$ ($N=2$, $N=3$) while $L \in (0,1)$ and with increasing of the value of the conformal parameter $N$ the mimic value of the coupling parameter $\omega_{\rm B}$ increases.

The studies of center-of-mass energies of collisions of two charged particles show that the increase of both conformal and spin parameters causes a decrease in the center-of-mass energy.

\section*{Acknowledgments}

This research is supported by the Uzbekistan Ministry for Innovative Development, Grants No. VA-FA-F-2-008 and No. MRB-AN-2019-29, the Innovation Program of the Shanghai Municipal Education Commission, Grant No.~2019-01-07-00-07-E00035, and the National Natural Science Foundation of China (NSFC), Grant No.~11973019. B.N. also acknowledges support from the China Scholarship Council (CSC), grant No.~2018DFH009013. This research is partially supported by an Erasmus+ exchange Grant between SU and NUUz.  A.A. is supported by a postdoc fund through PIFI of the Chinese Academy of Sciences.

\begin{appendix}

\section{Electromagnetic field components }\label{app}
\begin{widetext}
\begin{eqnarray}
\label{er} %
E^{\hat{r}}&=&\frac{a B \sqrt{\Delta }}{2
   \sqrt{{\cal Q}{\cal R}} \Sigma ^{5/2} \left(L^2+\Sigma \right)} \left[-40 a^2 L^2 M^2 r^3 \sin ^2\theta (\cos 2 \theta+3)+\Sigma ^2 \left(-M \cos 2 \theta \left(a^2 \left(2 r (2 M+r)-L^2\right)\right.\right.\right.
\\\nonumber   
&&  \left.\left.\left. -3 L^2 r^2+2 r^4\right)+3 a^2 L^2
   M+16 a^2 L^2 r-a^2 M^2 r \cos 4 \theta +5 a^2 M^2 r-6 a^2 M r^2+L^2 M r^2+16 L^2 r^3-6 M r^4\right)\right.
\\\nonumber   
&& \left.  -M r \Sigma  \left(a^2 M \cos 4 \theta  \left(L^2-2 r^2\right)+2 \cos 2 \theta \left(a^2
   \left(2 L^2 M+17 L^2 r-4 M r^2\right)+9 L^2 r^3\right)-5 a^2 L^2 M+6 a^2 L^2 r \right.\right.
\\\nonumber   
&&  \left.\left. +10 a^2 M r^2+22 L^2 r^3\right)+M \Sigma ^3 \left(\left(a^2+3 r^2\right) \cos 2 \theta+3 a^2+r^2\right)\right]\\\nonumber\\
\label{e2}  E^{\hat \theta}&=&  -\frac{a B \sin 2 \theta }{2 \sqrt{{\cal Q} {\cal R}} \Sigma ^{5/2} \left(L^2+\Sigma \right)} \left[-20 a^4 L^2 M^2 r^2 \sin ^2\theta (\cos 2 \theta +3)+a^2 M r \Sigma  \left(-\cos 2 \theta  \left(a^2 \left(17 L^2-4 M r\right)\right.\right.\right.
\\\nonumber
&&\left.\left.\left. +3 L^2 r (2 M+3 r)\right)-3 a^2
   L^2+a^2 M r \cos 4 \theta -5 a^2 M r-2 L^2 M r-11 L^2 r^2\right)+4 M r \Sigma ^3 \left(a^2+r^2\right)\right.
\\\nonumber   
&&\left.   +\Sigma ^2 \left(a^4 \left(8 L^2-3 M r\right)+a^2 r \left(4 L^2 M+8 L^2 r-2 M^2 r-3 M
   r^2\right)-a^2 M r \cos 2 \theta  \left(a^2+r (6 M+r)\right)+4 L^2 M r^3\right)\right]
\\\nonumber\\
\label{m1} B^{\hat r} &=&\frac{B}{2 \sqrt{{\cal R}} \Sigma ^2 \left(L^2+\Sigma \right)} \Big[\frac{1}{2} a^4 M r \sin \theta  \left(\Sigma  \sin 4 \theta +10 L^2 \sin \theta  \cos 3 \theta \right)+8 a^2 L^2 M r \Sigma  \cos ^3\theta \Big.
\\\nonumber
&& \Big.  -\cos \theta \left(2 \Sigma ^2
   \left(a^2 \left(L^2-2 M r+\Sigma \right)+r^2 \left(L^2+\Sigma \right)\right)-4 a^2 M r \Sigma ^2 \cos 2 \theta \right.\Big.
\\\nonumber   
&& \Big.\left.  +\sin ^2\theta  \left(a^4 \left(-35 L^2 M r+8 L^2 \Sigma -6 M r \Sigma \right)+8
   a^2 L^2 r^2 \Sigma \right)\right)\Big]\\\nonumber\\
\label{bt} B^{\hat\theta} &=& \frac{B \sqrt{\Delta } \sin \theta }{\sqrt{{\cal R}} \Sigma ^2 \left(L^2+\Sigma \right)} \left[5 a^2 L^2 M r^2 (\cos 2 \theta +3)+\Sigma  \left(a^2 M \sin ^2\theta  \left(L^2-2 r^2\right)\right.\right.
\\\nonumber
&& \left.\left. -2 \left(a^2 \left(L^2 M+2 L^2 r-2 M r^2\right)+2 L^2
   r^3\right)\right)+\Sigma ^2 \left(a^2 M \sin ^2\theta -2 a^2 M+L^2 r\right)+r \Sigma ^3\right]
\end{eqnarray}
where we used the following notations $${\cal R}=2 a^2 M r \sin ^2\theta +\Sigma  \left(a^2+r^2\right)\ ,$$ and $${\cal Q}=\Sigma  \left(a^2+r^2\right)-M r \left(a^2 \cos 2 \theta +a^2+2 r^2\right)\ .$$
\begin{eqnarray}\label{e1}
E^{\hat{r}}&=&\frac{a B \sqrt{\Delta }}{\Sigma
   ^{5/2} \sqrt{{\cal Q} {\cal R}}} \left[8 L^2 r \left\{-\frac{2 a^2 M^2 r^2 \sin ^2\theta (\cos (2 \theta
   )+3)}{\Sigma }-M r \left(\left(2 a^2+r^2\right) \cos 2 \theta+r^2\right)+\Sigma 
   \left(a^2+r^2\right)\right\}\right.
\\\nonumber   
&&\left.   +M \Big(-4 a^2 M r^3 \sin ^2\theta (\cos 2 \theta+3)-r \Sigma 
   (\cos 2 \theta+3) \left(r \left(a^2+r^2\right)-2 a^2 M \sin ^2\theta\right)\Big.\right.
\\\nonumber   
&&\left.\Big.   +\frac{\Sigma ^2}{2}
    \left(\left(a^2+3 r^2\right) \cos 2 \theta+3 a^2+r^2\right)\Big)\right] \ ,
\\\nonumber
\\
E^{\hat{\theta}}&=&\frac{a B \sqrt{\Delta }\sin 2 \theta}{\Sigma ^{5/2} \sqrt{{\cal Q} {\cal R}}} \left[4 a^2 L^2  \left\{\frac{2 a^2
   M^2 r^2 \sin ^2\theta (\cos 2 \theta +3)}{\Sigma }+M r \left(\left(2 a^2+r^2\right) \cos 2
   \theta +r^2\right)-\Sigma  \left(a^2+r^2\right)\right\}\right.
\\\nonumber   
&&\left.   +M r   \left(2 a^4 M r \sin ^2\theta (\cos 2 \theta
   +3)+\frac{1}{2} a^2 \Sigma  \left(\cos 2 \theta  \left(a^2+6 M r+r^2\right)+3 a^2+2 M r+3
   r^2\right)-2 \Sigma ^2 \left(a^2+r^2\right)\right)\right]\ ,
  \\
%
B^{\hat{r}}&=& \frac{B \cos \theta }{\sqrt{{\cal R}} \Sigma ^2} \Bigg[4 a^2 L^2 \sin ^2\theta
    \left\{a^2+r^2-\frac{a^2 M r (\cos 2 \theta +3)}{\Sigma }\right\}\Bigg.
\\\nonumber    
&&\Bigg.    -\left(a^4 M r \sin ^2\theta  (\cos 2 \theta +3)-4 a^2 M r \Sigma  \cos
   ^2\theta +\Sigma ^2 \left(a^2+r^2\right)\right)\Bigg]\ ,
   \\\nonumber
   \\\label{m2}
B^{\hat{\theta}}&=&   \frac{B \sqrt{\Delta } \sin \theta }{\sqrt{{\cal R}} \Sigma ^2} \left[4 L^2 r \left\{\frac{a^2 M r (3+\cos 2 \theta )}{\Sigma
   }-a^2-r^2\right\}+a^2 M \left(r^2-\frac{\Sigma}{2}\right)(3+\cos 2 \theta )+r\Sigma ^2\right]
\end{eqnarray}

Here $\wedge$
(hat) stands for orthonormal components of the electric and
magnetic fields.

\end{widetext}
\end{appendix}

\end{document}